# PrivPetal: Relational Data Synthesis via Permutation Relations

Kuntai Cai
National University of Singapore
Singapore, Singapore
caikt@comp.nus.edu.sg

Xiaokui Xiao
National University of Singapore
Singapore, Singapore
xkxiao@nus.edu.sg

Yin Yang
Hamad bin Khalifa University
Al Rayyan, Qatar
yyang@hbku.edu.qa

## Abstract

Releasing relational databases while preserving privacy is an important research problem with numerous applications. A canonical approach is to generate synthetic data under differential privacy (DP), which provides a strong, rigorous privacy guarantee. The problem is particularly challenging when the data involve not only entities (e.g., represented by records in tables) but also relationships (represented by foreign-key references), since if we generate random records for each entity independently, the resulting synthetic data usually fail to exhibit realistic relationships. The current state of the art, PrivLava, addresses this issue by generating random join key attributes through a sophisticated expectation-maximization (EM) algorithm. This method, however, is rather costly in terms of privacy budget consumption, due to the numerous EM iterations needed to retain high data utility. Consequently, the privacy cost of PrivLava can be prohibitive for some real-world scenarios.

We observe that the utility of the synthetic data is inherently sensitive to the join keys: changing the primary key of a record $t$, for example, causes $t$ to join with a completely different set of partner records, which may lead to a significant distribution shift of the join result. Consequently, join keys need to be kept highly accurate, meaning that enforcing DP on them inevitably incurs a high privacy cost. In this paper, we explore a different direction: synthesizing a *flattened* relation and subsequently decomposing it down to base relations, which eliminates the need to generate join keys. Realizing this idea is challenging, since naively flattening a relational schema leads to a rather high-dimensional table, which is hard to synthesize accurately with differential privacy.

We present a sophisticated PrivPetal approach that addresses the above issues via a novel concept: *permutation relation*, which is constructed as a surrogate to synthesize the flattened relation, avoiding the generation of a high-dimensional relation directly. The synthesis is done using a refined Markov random field mechanism, backed by fine-grained privacy analysis. Extensive experiments using multiple real datasets and the TPC-H benchmark demonstrate that PrivPetal significantly outperforms existing methods in terms of aggregate query accuracy on the synthetic data.

## 1 Introduction

Releasing relational databases while preserving privacy is an important and challenging research problem with a wide range of applications, e.g., in fields such as finance and healthcare. A common methodology is to generate synthetic data under differential privacy (DP) [11], which provides strong, rigorous privacy guarantees. Specifically, DP ensures that given the released synthetic data, the adversary cannot derive with high confidence (controlled by a parameter called the *privacy budget*) the presence or absence of a given record in the original database. Unlike other DP-based solutions, which release specific statistical information about the underlying data, differentially private synthesis outputs a complete database that retains the statistical properties of the original data, which can be used in downstream applications as if it were the original database. This offers significant convenience for the user; hence, the problem has attracted much research attention in recent years, e.g., [2, 5–7, 9, 10, 12, 17, 18, 20, 25, 27, 29–31, 37, 39, 41, 42, 44].

Earlier work primarily focused on synthesizing *single-relation* databases with DP. In practice, however, a relational database must capture not only entities (represented by records in a table), but also relationships (represented by foreign-key references across multiple tables). If we directly use existing solutions to generate random records for each relation, the result would likely fail to preserve the true relationships between entities, particularly the cross-table correlations induced by foreign keys. For example, consider the census database illustrated in Table 1, which includes an individual table $R_I$ and a household table $R_H$. $R_I$ contains a foreign key H-ID referencing $R_H$ to indicate each individual's corresponding household. To effectively synthesize this database, it is necessary to preserve three types of correlations in the synthetic data:

- **Inter-attribute correlations** among attributes in the same relation, e.g., how individuals' ages correlate with their education levels.
- **Intra-group correlations** among individual tuples with the same foreign key, e.g., what types of individuals (as indicated by their attributes in $R_I$) tend to reside in the same household.

**Table 1: Example data in a relational database.**

(a) Individual relation $R_I$.

| AGE | EMP | EDU  | MAR | H-ID |
|-----|-----|------|-----|------|
| 40  | Yes | Mid  | Yes | 1    |
| 35  | Yes | High | Yes | 1    |
| 12  | No  | Low  | No  | 1    |
| 10  | No  | Low  | No  | 1    |
| 55  | No  | High | Yes | 2    |
| 60  | No  | Mid  | Yes | 2    |
| 25  | Yes | Mid  | No  | 2    |
| 30  | Yes | High | Yes | 3    |
| 25  | Yes | High | Yes | 3    |

(b) Household relation $R_H$.

| H-ID | OWN |
|------|-----|
| 1    | No  |
| 2    | No  |
| 3    | Yes |

Inter-relational ↔

Intra-group

Inter-attribute

- **Inter-relational correlations** between tuples in $R_I$ and $R_H$, e.g., what kinds of individuals are likely to own their homes (indicated by the OWN attribute in $R_H$).

Previous single-relation data synthesis methods have demonstrated the capability to capture inter-attribute correlations. However, extending these methods to multi-relation cases is non-trivial. For example, if we synthesize $R_I$ and $R_H$ separately using a single-relation method and then randomly associate the tuples using H-ID, each individual tuple will have exactly the same probability of being associated with each household tuple, which breaks the inter-relational correlations between the individual and household tables. Similarly, intra-group correlations are also lost, as individual tuples are grouped into households randomly.

The task of generating appropriate join keys for synthetic relations is particularly challenging, as foreign keys (e.g., H-ID of the above example) follow a rather complex distribution that depends on attributes in both the referencing table (i.e., $R_I$) and the referenced table ($R_H$). For instance, to generate H-ID accurately, we must address two key aspects: (i) tracking which individual tuples have been assigned to specific H-ID values, and (ii) ensuring that the assigned household tuples align with the relationships observed in the original data. These challenges become increasingly difficult as the number of referencing tuples and attributes increases.

Given the difficulties of generating foreign keys to accurately link synthetic tuples across relations, an alternative solution is to synthesize the join of $R_I$ and $R_H$ directly. Two possible strategies can be considered. The first involves removing H-ID, synthesizing the join using single-relation methods, and then randomly grouping synthetic individual tuples into households. While this method partially retains inter-relational correlations, it fails to capture intra-group correlations. Specifically, without H-ID, it is impossible to accurately infer the coexistence of individuals within the same household. The second strategy preserves H-ID and synthesizes the join directly. However, this introduces significant challenges due to the large domain size of H-ID. Existing single-relation synthesis methods (e.g. [5, 6, 9, 17, 29, 30, 39, 41]) are not well-equipped to handle such large domains, as they typically require querying the distribution of attribute values, and the noise of DP injected into H-ID's distribution can severely compromise its statistical properties, resulting in poor-quality synthetic data.

In the DP data synthesis literature, a notable solution for synthesizing multi-relation data is PrivLava [7], which works by introducing *type attributes* as a means of modeling intra-group and inter-relational correlations. In the above example, PrivLava first clusters households into different types, represented by a new TYPE column in $R_H$ (elaborated later in Section 2.3), based on attributes of its composing individuals (e.g., one type might be households with two employed, married and highly educated individuals). Then, PrivLava synthesizes $R_H$ along with the calculated TYPE attribute. After that, the algorithm generates $R_I$ by creating random individuals corresponding to the types of households.

The type attributes in PrivLava are essentially surrogates for the join keys, which mitigate the difficulty of generating effective keys directly, as the type attributes have a coarser granularity.

However, the underlying challenges persist: the type attributes still follow a complex distribution, in that the types (i.e., clustering results) depend not only on the corresponding individuals but also on other types. This self-dependency compels PrivLava to use an iterative approach to refine and determine the types, and each iteration inevitably queries information, resulting in a high overall privacy cost. The problem is exacerbated when the data is highly diverse, as more types are required to model the data accurately. Indeed, as we show later in Section 6.4, PrivLava performs poorly when the privacy budget is restricted. This observation further reinforces that there is considerable room for improvement over PrivLava.

**Our contribution.** In this paper, we present PrivPetal (differentially private data synthesis via permutation relations), a novel solution for synthesizing multi-relation databases with foreign key references under DP. PrivPetal avoids the hard problem of generating join keys under DP altogether, and instead synthesizes a *flattened relation* that merges data from both the referencing and the referenced tables. Because the flattened relation has a very high dimensionality and, thus, is hard to synthesize directly under DP, PrivPetal introduces a novel concept of *permutation relations* (PRs), which enumerate all possible permutations of data tuples in the referencing relation (e.g., $R_I$).

Specifically, PrivPetal (i) computes the PR from the flattened relation, (ii) queries marginal distributions from the PR, (iii) constructs graphical models based on these marginals, (iv) uses the constructed models to synthesize the flattened relation, and finally (v) decomposes the synthetic flattened relation to obtain the base relations. As the marginals are built upon the original flattened data encompassing both tables, the generated synthetic database effectively preserves inter-attribute, intra-group, and inter-relations correlations. Here, step (iii) is performed using an adapted PrivMRF [6] algorithm, with rigorous, fine-grained privacy analysis that exploits properties of the permutation relation. PrivPetal naturally extends to the general case with multiple tables and foreign keys, by processing each pair of relations linked by a foreign key.

We evaluate PrivPetal empirically on real-world datasets and the TPC-H benchmark, against state-of-the-art DP synthesis methods, including PrivLava and single-relation methods. The results demonstrate that PrivPetal significantly and consistently outperforms all its competitors in terms of data utility measured by the accuracy of join-aggregate queries on the synthetic database. We have open-sourced the code of PrivPetal at https://github.com/caicre/PrivPetal.

## 2 Preliminaries
### 2.1 Problem Setting

Let $R$ be a relation and $\mathcal{A}$ denote the set of all attributes in $R$, excluding primary and foreign keys. We assume that each attribute $A \in \mathcal{A}$ takes values from a discrete domain $\mathcal{X}_A$. This aligns with previous data synthesis methods [5–7, 9, 17, 24], and does not compromise generality, since continuous attributes can be discretized in a preprocessing step.

For any tuple $t \in R$, the value of $t$ in column $A$ is denoted as $t[A] \in \mathcal{X}_A$. Furthermore, for a set of attributes $S \subset \mathcal{A}$, we denote the domain of $S$ by $\mathcal{X}_S$, and the values of $t$ for the attributes in $S$ by $t[S] \in \mathcal{X}_S$. A database is a collection of relations $\mathcal{R} = \{R_0, R_1, R_2, \ldots\}$. For any two relations $R_i, R_j \in \mathcal{R}$ where $R_i$ contains a foreign key referencing the primary key of $R_j$, we say $R_i$ *refers to* $R_j$, and denote the foreign key as $\text{FK}(R_i, R_j)$. A tuple $t_i \in R_i$ *refers to* a tuple $t_j \in R_j$, if the foreign key of $t_i$ equals the primary key of $t_j$. Further, a tuple $t$ *depends on* $t_j$, if $t$ either refers to $t_j$, or refers to another tuple that depends on $t_j$, i.e., $t_i$ connects to $t_j$ via a chain of foreign-key-to-primary-key relationships. Similarly, a relation $R_i$ *depends on* $R_j$, if $R_i$ either directly refers to $R_j$, or refers to another relation depending on $R_j$.

Without loss of generality, we assume that $R_0$ (called the *primary private relation*) contains sensitive information that must be protected under differential privacy (DP). All relations that depend on $R_0$ are called *secondary private relations*, which are also considered sensitive. Relations not included in these categories are considered public, and can be released freely without compromising individuals' privacy. These assumptions are consistent with previous work, e.g., in PrivLava [7]. We call a foreign key $\text{FK}(R_i, R_j)$ a *private foreign key* if $R_i$ is a private relation. Following PrivLava, we assume that the private foreign key references form a directed acyclic graph (DAG), i.e., a private $R$ is not allowed to depend on itself.

Given a database $\mathcal{R}$ and any tuple in its primary relation $t_0 \in R_0$, we can obtain a *neighbor database* $\mathcal{R}'$ of $\mathcal{R}$ by removing $t_0$ and all tuples in $\mathcal{R}$ dependent on $t_0$. Then, DP is defined as follows.

*Definition 2.1 (Differential Privacy [11]).* Let $F$ be an algorithm that takes a database as input. $F$ satisfies $(\epsilon, \delta)$-differential privacy, if and only if for any two neighboring databases $\mathcal{R}$ and $\mathcal{R}'$ and any possible set $O$ of outputs from $F$,

$$\Pr[F(\mathcal{R}) \in O] \leq e^{\epsilon} \cdot \Pr[F(\mathcal{R}') \in O] + \delta.$$

In the literature, $\epsilon$ is commonly referred to as the *privacy budget*, which indicates the level of privacy protection (a smaller $\epsilon$ corresponds to a stronger privacy guarantee), and $\delta$ is usually fixed to a small constant not exceeding the inverse of the dataset size. In our setting, DP ensures that the presence or absence of any single tuple $t_0$ in the primary private relation $R_0$, as well as all other tuples that depend on $t_0$, do not affect the outcome of algorithm $F$ significantly. Therefore, the adversary cannot confidently determine whether $t_0$ is included in $R_0$ based on the output of $F$, even with extensive background knowledge of other tuples and relationships in $\mathcal{R}$. Our objective is then to synthesize a relational database with high utility, while satisfying DP. The specific definition of data utility depends on the application; intuitively, it should incorporate inter-attribute, intra-group, and inter-relational correlations described in Section 1.

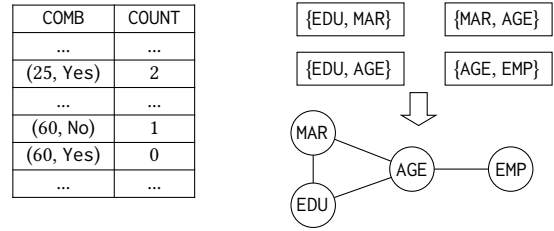

(a) Marginal on {AGE, EMP}. (b) From marginals to graphical model.
**Figure 1: Marginal and graphical model.**

In Section 6, we present example utility measures based on the accuracy of join-aggregate query results on the synthetic data compared to the result of the same query on the original real data.

### 2.2 Marginal-Based Data Synthesis

State-of-the-art DP synthesis methods [5, 6, 9, 29, 30, 41, 42] typically focus on single-relation databases, and rely on marginal distributions (hereafter referred to as *marginals*). We illustrate these methods using $R_I$ from our running example in Table 1a; here, we ignore the foreign key H-ID of $R_I$ as we are discussing a single-relation setting. A marginal is a contingency table that counts the occurrences of attribute value combinations. For example, Figure 1a presents the marginal on the attribute set {AGE, EMP}. The combination (60, No) occurs once in $R_I$; hence, there is a record in the marginal with count 1 for this attribute combination. Similarly, the combination (25, Yes) appears twice, hence the corresponding count of 2 in the marginal. Clearly, a marginal represents the joint data distribution for a subset of attributes, capturing their correlations.

A typical DP-compliant single-relation data synthesis method follows these steps: (i) select marginals with correlated attributes, (ii) inject noise into the marginal counts to preserve DP, (iii) construct a graphical model [36] from the noisy marginals, and (iv) use this graphical model to sample tuples and synthesize $R_I$. In particular, a graphical model is a statistical tool for describing probability distributions. As exemplified by Figure 1b, it can be constructed from the marginals on {EDU, MAR}, {EDU, AGE}, {MAR, AGE}, {AGE, EMP}, and produces an overall distribution $\Pr[\text{AGE, EMP, EDU, MAR}]$, which approximates the original distribution of tuples in $R_I$, as both share the same marginals included in the graphical model. Thus, sampling tuples from this model typically results in a synthetic $R_I$ with high utility, in the sense that the synthetic data preserve the real data's inter-attribute correlations for attributes involved in the marginals.

In particular, the sampling process utilizes the conditional distributions derived from the graphical model. In the example shown in Figure 1a, a typical single-relation data synthesis method first computes $\Pr[\text{EDU}]$, $\Pr[\text{MAR} \mid \text{EDU}]$, $\Pr[\text{AGE} \mid \text{MAR, EDU}]$, and $\Pr[\text{EMP} \mid \text{AGE}]$ from the graphical model using standard algorithms. Then, it samples attributes sequentially, with each attribute conditioned on previously sampled attributes, e.g.,

(1) Sample $\text{EDU} = x_1$ with $\Pr[\text{EDU} = x_1]$.
(2) Sample $\text{MAR} = x_2$ with $\Pr[\text{MAR} = x_2 \mid \text{EDU} = x_1]$.
(3) Sample $\text{AGE} = x_3$ with $\Pr[\text{AGE} = x_3 \mid \text{MAR} = x_2, \text{EDU} = x_1]$.
(4) Sample $\text{EMP} = x_4$ with $\Pr[\text{EMP} = x_4 \mid \text{AGE} = x_3]$.

This produces a synthetic tuple with $\text{EDU} = x_1$, $\text{MAR} = x_2$, $\text{AGE} = x_3$, and $\text{EMP} = x_4$. The sampling is repeated until a sufficient number of tuples are generated to form the synthetic $R_I$.

**Table 2: PrivLava example.**

(a) Augmented $R_I$.

| AGE | EMP | EDU  | MAR | H-ID | TYPE |
|-----|-----|------|-----|------|------|
| 40  | Yes | Mid  | Yes | 1    |      |
| 35  | Yes | High | Yes | 1    | A    |
| 12  | No  | Low  | No  | 1    |      |
| 10  | No  | Low  | No  | 1    |      |
| 55  | No  | High | Yes | 2    |      |
| 60  | No  | Mid  | Yes | 2    | B    |
| 25  | Yes | Mid  | No  | 2    |      |
| 30  | Yes | High | Yes | 3    | C    |
| 25  | Yes | High | Yes | 3    |      |

(b) Augmented $R_H$.

| TYPE | H-ID | OWN |
|------|------|-----|
| A    | 1    | No  |
| B    | 2    | No  |
| C    | 3    | Yes |

**Table 3: Flattened relation for the running example.**

| $I_1$.AGE | $I_1$.EMP | $I_2$.AGE | $I_2$.EMP | ... | $I_N$.AGE | $I_N$.EMP | $H$.OWN | $H$.SIZE |
|-----------|-----------|-----------|-----------|-----|-----------|-----------|---------|----------|
| 40        | Yes       | 35        | Yes       | ... | 10        | No        | No      | 4        |
| 55        | No        | 60        | No        | ... | NULL      | NULL      | No      | 3        |
| 30        | Yes       | 25        | Yes       | ... | NULL      | NULL      | Yes     | 2        |

⎵ Individual 1 ⎵ ⎵ Individual 2 ⎵ ... ⎵ Individual N ⎵ ⎵ Household ⎵

## 3 Main Ideas of PrivPetal

Recall from Section 1 and Section 2.3 that generating synthetic join keys under DP that effectively capture intra-group and inter-relation correlations is a difficult problem, as correct keys are vital for the output data utility, and yet the keys follow complex distributions that are expensive to approximate in terms of privacy cost. Motivated by this, we explore a radically different direction that avoids generating join keys altogether: synthesizing a *flattened relation* (FR) that merges multiple tables under DP, and subsequently decomposing the FR to obtain the base tables.

Our PrivPetal algorithm builds on this idea, tackling several non-trivial challenges along the way, including the high dimensionality of the FR and the complexities of its synthesis. In the following, Section 3.1 presents the concepts of FR and *permutation relation* (PR), a surrogate for synthesizing the high-dimensional FR. Section 3.2 introduces *normalized permutation marginals* (NPMs), derived from PRs, to capture correlations in the FR. Section 3.3 explains how PrivPetal creates a synthetic FR through NPMs.

### 3.1 Flattened and Permutation Relations

**Flattened relation.** Consider again our running example, with primary private relation $R_H$ and secondary private relation $R_I$. Table 3 illustrates the FR, in which each tuple represents a household, formed by concatenating a tuple in $R_H$ with all its join partners in $R_I$, i.e., individual members of the household (for brevity, only AGE and EMP attributes from $R_I$ are shown). To distinguish attributes from different tables, we prefix them with $H$ (for the household tuple), $I_1$ (the first join partner), $I_2$ (second individual), ..., $I_N$, where $N$ is the maximum number of individuals in any household ($N = 4$ in this example). The FR also includes a derived attribute $H$.SIZE, which records the number of individuals in each household.

Formally, consider two relations $R_I$ and $R_H$, with a foreign key $A_{FK}$ in $R_I$ referencing the primary key $A_{ID}$ of $R_H$. We denote the attribute sets of $R_I$ and $R_H$ by $\mathcal{A}_I = \{I.A_1, I.A_2, \ldots\}$ and $\mathcal{A}_H = \{H.A_1, H.A_2, \ldots\}$, respectively, excluding the join keys. We assume that $R_H$ is augmented to include a special attribute $H.A_{size} \in \mathcal{A}_H$ (e.g., the derived $H$.SIZE in Table 3), which represents the number of tuples in $R_I$ referencing each household tuple.

To construct the FR, we group tuples in $R_I$ by their foreign key $A_{FK}$. A *flattened tuple* in the FR is then obtained by concatenating all $R_I$ tuples in a group and the associated $R_H$ tuple. For instance, in Table 1, the top four individuals in $R_I$ (with the same foreign key H-ID = 1) are concatenated with the first household in $R_H$, to form the first row in Table 3. Given the maximum group size $N$, we pad NULLs in a flattened tuple, if its corresponding group contains less than $N$ tuples from $R_I$ (e.g., in the second and third rows of Table 3). Then, the FR of $R_I$ and $R_H$, denoted by $F(R_I, R_H)$, consists of all the flattened tuples produced this way.

As in Table 3, in the FR, we prefix the attributes that come from the $i$-th member of the household with $I_i$. We denote the attribute

### 2.3 PrivLava

To our knowledge, PrivLava [7] is currently the only solution optimized for synthesizing multiple-relation databases under DP, which generates join keys guided by *type attributes* learned from the data. Table 2a illustrates PrivLava for our running example, assuming that $R_H$ is the primary private relation (explained in Section 2.1). First of all, PrivLava clusters the tuples in the referenced table (i.e., $R_H$) into different *types*, and adds them as a derived attribute in $R_H$ (TYPE in Table 2b). Note that the computation of record types takes into account attribute values of the join partners in the referencing table ($R_I$). For instance, TYPE-A can be households with a middle-aged employed couple and two children, and TYPE-B can be those with an elderly unemployed couple and a young individual, etc.

Then, PrivLava applies a single-relation method to synthesize the augmented $R_H$, which includes the derived TYPE attribute. After that, PrivLava augments the referencing relation $R_I$ by adding the TYPE of each individual's corresponding household, and then proceeds to synthesize $R_I$ by creating a group of random individuals for each synthetic household, i.e., a tuple in the generated $R_H$. Specifically, the algorithm first constructs a graphical model to approximate the tuple distribution in the original $R_I$, including the TYPE attribute. Then, the method samples attributes for each individual, conditioned on TYPE, using the sampling process described in Section 2.2. This approach ensures that the synthetic individuals are consistent with their respective TYPE attribute, which helps preserve intra-group and inter-relational correlations.

As mentioned in Section 1, PrivLava still suffers from the complex distribution that the new TYPE attributes follow. In particular, they are clustering results depending not only on the corresponding individuals but also on other types. In PrivLava, the computation of TYPE is based on an iterative expectation-maximization algorithm. Each iteration involves private information, and, thus, incurs a privacy cost under DP. To obtain accurate record types, PrivLava often needs many iterations, leading to a high overall privacy cost, or, equivalent, suboptimal result utility under a given limited privacy budget $\epsilon$. Further, when the underlying data exhibit high diversity, the benefit of clustering records into coarser types diminishes, as the TYPE attribute needs to have a large number of distinct values to properly model various clusters, making it even harder to synthesize accurately under DP, leading to degraded quality of the output synthetic data.

set of the $i$-th individual by $\mathcal{A}_{I_i} = \{I_i.A_1, I_i.A_2, \ldots\}$. Finally, for each $s \leq N$, let $n_s$ be the number of flattened tuples of group size $s$ in $F(R_I, R_H)$, and $n = \sum_{s \leq N} n_s$ be the total number of flattened tuples in the FR. For flattened tuples with group size $s$, we denote its attribute set by $\mathcal{A}_s = \mathcal{A}_H \cup \mathcal{A}_{I_1} \cup \cdots \cup \mathcal{A}_{I_s}$. Then, $\mathcal{A}_N$ denotes the attribute set of $F(R_I, R_H)$.

Observe that an FR typically has a high dimensionality, since each flattened tuple is produced by concatenating $N$ records from $R_I$ (including padded ones) and one from $R_H$. This poses a serious problem when applying DP-compliant data synthesis: recall from Section 2.2 that a relation is typically synthesized through marginals, each corresponding to an attribute set. Clearly, the number of possible marginals grows exponentially with the data dimensionality, which is prohibitively expensive for a table with numerous columns such as an FR, as the computation of each marginal incurs a privacy cost under DP. Even if we restrict the number of attributes in a marginal, the number of possible marginals still grows rapidly with $N$, which can be impractical with a larger $N$.

To tackle this problem, we introduce a novel concept of *permutation relation* (PR), which approximates an FR using possible combinations of individual tuples, explained below.

**Permutation relation.** As mentioned above, directly synthesizing the FR is challenging due to its high dimensionality. In addition, two more subtle yet critical issues arise when attempting to synthesize the FR. First, the FR introduces an *artificial order* in each flattened tuple, for its $R_I$ components. For instance, in Table 3, the first flattened tuple contains the attributes of 4 individuals in $R_I$ who belong to the same household, labeled as $I_1, I_2, I_3$, and $I_4$. The order of these individuals is completely arbitrary; it affects the marginal-based data synthesis process described in Section 2.2. For instance, a marginal with the attributes $\{I_1.\text{AGE}, H.\text{OWN}\}$ clearly depends on *who* in this household is designated as $I_1$, which is an arbitrary decision. Preserving such a marginal in the synthetic data is rather pointless.

Second, the number of marginals grows exponentially with the cardinality of its attribute set. In practice, this forces data synthesis solutions to limit the number of attributes in a marginal. Clearly, this also limits the number of base tuples whose attributes can be included in a marginal, which implies that a flattened tuple only needs to include a much smaller number of base records.

The above insights lead to the concept of permutation relation (PR), in which (i) each tuple contains one record from the referenced table $R_H$ and a small number ($\ll N$) of records from the referencing table $R_I$, and (ii) the records from $R_I$ are permuted, forming multiple tuples in the PR. Table 4 shows a portion of the PR for our running example, for households with 3 members, i.e., $H.\text{SIZE} = 3$. Each tuple contains the attributes of a household (e.g., the second row in Table 1b), as well as 2 individuals (out of the 3 members in the household). The PR includes all permutations of the 2 individuals, e.g., the first two rows in Table 4 differ only by the order of the two individuals. Attributes are prefixed with $I_a, I_b, \ldots$ instead of $I_1, I_2, \ldots$ to indicate that they are not tied to specific positions.

PrivPetal synthesizes tuples in the FR for each group size separately. Consider synthesizing flattened tuples of group size $s$, and permutations involving at most $o$ individuals. The *order-$o$ permutation relation* for group size $s$, denoted as $P_{s,o}(R_I, R_H)$, enumerates all possible $o$-individual permutations from flattened tuples of group size $s$. For example, Table 4 shows $P_{3,2}(R_I, R_H)$. In PrivPetal, $o$ is a system parameter; thus, we omit $o$ and use $P_s(R_I, R_H)$ to denote the PR with group size $s$, and use $\mathcal{A}_{P_s}$ to denote the attribute set of $P_s(R_I, R_H)$, excluding primary and foreign keys.

**Table 4: Permutation relation for group size 3.**

| $I_a$.AGE | $I_a$.EMP | $I_b$.AGE | $I_b$.EMP | $H$.OWN | $H$.SIZE |
|---|---|---|---|---|---|
| 55 | No | 60 | No | No | 3 |
| 60 | No | 55 | No | No | 3 |
| 55 | No | 25 | Yes | No | 3 |
| 25 | Yes | 55 | No | No | 3 |
| … | … | … | … | … | … |
| Individual a | | Individual b | | Household | |

### 3.2 Normalized Permutation Marginals

Recall from Section 2.2 that DP-compliant data synthesis methods typically use marginals to preserve statistical properties of the underlying data. Formally, given a relation $R$ and a subset $S$ of its attributes, the marginal of $R$ on $S$, denoted by $M_S$, is a contingency table such that for each specific attribute value combination $x \in \mathcal{X}_S$, the cell $M_S[x]$ counts the occurrences of $x$ in the column set $S$, as explained in Section 2.2. Marginals computed from a permutation relation $P_s(R_I, R_H)$ are referred to as *permutation marginals*, which can be utilized to capture different types of correlations in the FR. For instance, in the PR in Table 4 derived from the FR in Table 3:

- The permutation marginal on $\{H.\text{OWN}, I_a.\text{EMP}\}$ captures the inter-relational correlation for attribute sets $\{H.\text{OWN}, I_1.\text{EMP}\}$, $\{H.\text{OWN}, I_2.\text{EMP}\}$, etc., in the FR.
- The permutation marginal on $\{I_a.\text{EMP}, I_b.\text{EMP}\}$ captures the intra-group correlation for attribute sets $\{I_1.\text{EMP}, I_2.\text{EMP}\}$, $\{I_1.\text{EMP}, I_3.\text{EMP}\}$, $\{I_2.\text{EMP}, I_3.\text{EMP}\}$, etc., in the FR.
- The permutation marginal on $\{I_a.\text{AGE}, I_a.\text{EMP}\}$ captures the inter-attribute correlation for attribute sets $\{I_1.\text{AGE}, I_1.\text{EMP}\}$, $\{I_2.\text{AGE}, I_2.\text{EMP}\}$, etc., in the FR.

Formally, let $m$ be an arbitrary injective mapping from any letter $x$ to an integer. By replacing each individual identifier $I_x$ in $S$ with $I_{m(x)}$, we obtain the corresponding attribute set $S' \subset \mathcal{A}_N$ in the FR. Let $S \to S'$ denote the relationship between $S$ and $S'$.

**Normalization.** Unlike ordinary relations, a PR contains rows with redundant information, causing the counts in the permutation marginals to be larger compared to those derived directly from the original data, i.e., from the FR. For example, consider the attribute set $S = \{I_a.\text{EMP}, I_b.\text{EMP}\}$ in the running example. The first two rows in the PR shown in Table 4 both contribute to the count for cell (No, No) in marginal $M_S$. However, these two rows should only be counted once, as they originate from the same row in the FR (i.e., second row in Table 3), with different permutations of the two individuals. This motivates us to introduce *normalized permutation marginals* (NPMs), which are normalized such that the total in each NPM equals $n_s$, i.e., the number of flattened tuples of group size $s$ in the FR. Formally, for any attribute set $S \subset \mathcal{A}_{P_s}$, the NPM on $S$ for group size $s$ is denoted by $M_{S,s}$. For any attribute value combination $x \in \mathcal{X}_S$, the cell $M_{S,s}[x]$ counts the occurrences of $x$ in columns $S$ in $P_s(R_I, R_H)$, multiplied by a normalizing factor $\frac{1}{W_s}$, where

$$W_s = \begin{cases} \frac{s!}{(s-o)!}, & \text{if } s \geq o, \\ s!, & \text{if } s < o, \end{cases} \quad (1)$$

is the number of permutation tuples generated by each flattened tuple, and $o$ is the order of the PR which is a system parameter, as described in the previous subsection. After normalization, the contributions of each flattened tuple sum up to 1, which bounds the sensitivity of each flattened tuple, which is important to limit the privacy cost of producing the NPM under DP.

**Equivalence to FR marginals.** Intuitively, if we permute all individual tuples within each flattened tuple in the FR, then $M_{S,s}$ is equivalent to the marginals counted directly from the FR. By this equivalence, we can construct graphical models from NPMs to approximate the distribution of flattened tuples of each group size $s$, while preserving the underlying correlations. Formally, we have the following result (proof can be found in Appendix A).

THEOREM 3.1 (APPLICABILITY OF NPMs). *For any group size $s \leq N$, let $F_s(R_I, R_H) \subseteq F(R_I, R_H)$ be the subset of flattened tuples of group size $s$. Let $\mathbb{F}_s(R_I, R_H)$ be the collection of all the tuples, each given by removing* NULLs *and then permuting the individual tuples in a flattened tuple in $F_s(R_I, R_H)$. For any attribute sets $S \subset \mathcal{A}_{P_s}$ and $S' \subset \mathcal{A}_s$ such that $S \to S'$, and any $x \in \mathcal{X}_S$:*

$$M_{S,s}[x] = \frac{1}{s!} \mathbb{M}_{S',s}[x],$$

*where $\mathbb{M}_{S',s}$ is the marginal counted from $\mathbb{F}_s(R_I, R_H)$ on $S'$.*

By definition, an NPM can be rolled up to obtain another NPM with a smaller attribute set. Specifically, for any attribute set $S' \subset S$, we can obtain $M_{S',s}$ by summing over $M_{S,s}$, i.e., $\forall x' \in \mathcal{X}_{S'}$, we have:

$$M_{S',s}[x'] = \sum_{x \in \mathcal{X}_S, x[S']=x'} M_{S,s}[x]. \quad (2)$$

This property is particularly useful when $M_{S,s}$ is queried from data, but only a subset $S'$ is required for constructing graphical models. For example, we may have queried $M_{S,s}$ for $S = \{I_a.\text{AGE}, I_a.\text{EMP}, I_b.\text{AGE}\}$, while the NPM on $S' = \{I_a.\text{AGE}, I_a.\text{EMP}\}$ is required. In this case, we can roll up $M_{S,s}$ to $M_{S',s}$ and use it to construct the model. In the next subsection, we explain how PrivPetal builds graphical models and samples synthetic data using NPMs.

### 3.3 Data Synthesis via NPMs

Recall that the main idea of PrivPetal is to synthesize a flattened relation (FR) by processing flattened tuples with different group sizes $s$ separately, and subsequently decompose the synthetic FR into base relations as the final output data. As explained before, the FR has rather high dimensionality, meaning that it is infeasible to synthesize an FR, even with a state-of-the-art marginal-based single-relation solution, e.g., [6, 29]. Instead, we need a novel, scalable solution for generating a synthetic FR with NPMs.

Observe that the FR contains all the rows and columns of the referenced relation ($R_H$ in our running example), except for its primary key column. Therefore, this part of the FR can be obtained by running an existing single-relation solution, which produces a synthetic $R_H$. Next, PrivPetal generates the columns of the FR corresponding to each of the $N$ partners in a sequential manner, using graphical models computed from the NPMs. Unlike existing solutions that apply a single graphical model for the entire data synthesis process, PrivPetal builds multiple graphical models, one for each column in the FR corresponding to the $N$ join partners. As we explain soon, each graphical model deals with only a selected subset of the columns in the FR (i.e., only a few selective NPMs are involved in the construction of each graphical model), and an NPM can be re-used to estimate multiple marginals in these graphical models, which effectively limits the impact of the high dimensionality of the FR.

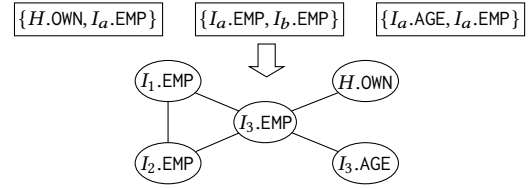

**Figure 2: Sample $I_3$.EMP using NPMs.**

We explain the idea of PrivPetal's attribute synthesis process through a concrete example. Consider the synthesis of $I_3$.EMP in the FR shown in Table 3. Suppose all other attributes have been synthesized. The graphical model for sampling $I_3$.EMP is depicted in Figure 2, which contains 5 attributes $I_1$.EMP, $I_2$.EMP, $I_3$.EMP, $I_3$.AGE, and $H$.OWN. PrivPetal constructs this graphical model to approximate $\Pr[H.\text{OWN}, I_1.\text{EMP}, I_2.\text{EMP}, I_3.\text{EMP}, I_3.\text{AGE}]$, conditioned on group size = 3, using three NPMs:

- The NPM on $\{H.\text{OWN}, I_a.\text{EMP}\}$ is used for $\{H.\text{OWN}, I_3.\text{EMP}\}$.
- The NPM on $\{I_a.\text{EMP}, I_b.\text{EMP}\}$ is used for $\{I_1.\text{EMP}, I_2.\text{EMP}\}$, $\{I_2.\text{EMP}, I_3.\text{EMP}\}$, and $\{I_1.\text{EMP}, I_3.\text{EMP}\}$.
- The NPM on $\{I_a.\text{AGE}, I_a.\text{EMP}\}$ is used for $\{I_3.\text{AGE}, I_3.\text{EMP}\}$.

Observe that one NPM can be used to approximate multiple marginals of the FR corresponding to different attribute groups, meaning that the number of NPMs that need to be computed can be far lower than the number of marginals involved in the graphical model, which significantly improves the efficiency of the algorithm, since computing each NPM (or marginal) incurs a privacy cost under DP.

PrivPetal then computes $\Pr[I_3.\text{EMP} \mid H.\text{OWN}, I_1.\text{EMP}, I_2.\text{EMP}, I_3.\text{AGE}]$ with this graphical model, and samples $I_3$.EMP conditioned on the other attributes. In general, for each individual attribute $I_i.A_j \in \mathcal{A}_N$ to be sampled and each possible group size $s \geq i$, PrivPetal constructs a graphical model to derive such a sampling probability, denoted as $p_s(I_i.A_j \mid C)$, where $C$ is the set of selected attributes correlated to $I_i.A_j$. Then, the values of $I_i.A_j$ for each tuple $t$ in the synthetic FR are sampled using the model corresponding to group size $t[H.A_{\text{size}}]$. If every attribute $I_i.A_j$ is sampled in this manner, the synthetic FR will mimic the original FR, and preserve the underlying correlations captured by NPMs.

The above description of the FR synthesis process leaves out a few important details, such as the choice of $C$, what NPMs are used for constructing graphical models, and how to build such a model under DP. We elaborate on them in the next section.

## 4 Detailed PrivPetal Algorithm

Sections 4.1- 4.3 focus on how PrivPetal synthesizes two relations linked by a foreign key, covering the overall flow (Section 4.1), the selection of relevant attributes from the FR (Section 4.2), and the construction of the graphical models (Section 4.3). Then, Section 4.4 extends PrivPetal to the general case with multiple private relations and foreign keys.

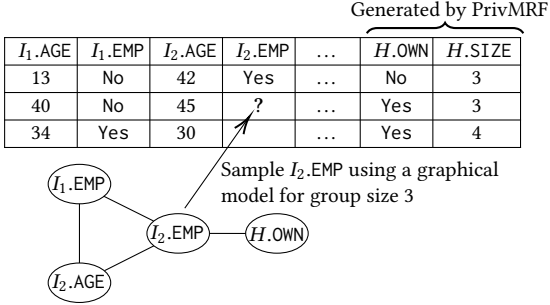

Figure 3: Attribute value synthesis in PrivPetal.

### 4.1 Solution Overview

Given a referencing relation $R_I$ and a referenced relation $R_H$, PrivPetal creates a synthetic version of the flattened relation $F(R_I, R_H)$, denoted as $\widetilde{F(R_I, R_H)}$. To do so, the solution first adapts a state-of-the-art single-relation method, PrivMRF [6], to synthesize $R_H$, which constitutes the columns of $\widetilde{F(R_I, R_H)}$ corresponding to $R_H$. Then, PrivPetal synthesizes each individual attribute $I_i.A_j$ sequentially to complete $\widetilde{F(R_I, R_H)}$, as demonstrated by Figure 3, with the following steps:

- **Selecting Correlated Attributes**: Identify a set $C$ of attributes that have already been synthesized, which are correlated to $I_i.A_j$.
- **Constructing Graphical Models**: For each group size $s$, construct a graphical model from NPMs to derive a distribution $p_s(C \cup \{I_i.A_j\})$, and compute $p_s(I_i.A_j \mid C)$ accordingly.
- **Sampling Attribute Values**: For each tuple $t \in \widetilde{F(R_I, R_H)}$, inspect its group size $s_t = t[H.A_{\text{size}}]$ and use the corresponding $p_{s_t}(I_i.A_j \mid C)$ to sample $t[I_i.A_j]$ based on $t[C]$.

Algorithm 1 overviews the PrivPetal algorithm, where the inputs consist of the two private relations $R_I$, $R_H$, as well as 4 scale parameters for injecting random noise required to satisfy DP. The values of these scale parameters are determined later in Section 5. In particular, we assume that the input relation $R_H$ is already augmented with the derived attribute $H.A_{\text{size}}$, which records the number of join partners in $R_I$ for each $R_H$ tuple, as described in Section 3.1.

The algorithm starts by invoking an adapted version of PrivMRF [6] (explained soon) to synthesize $R_H$ (Line 1), populating columns in the synthetic FR corresponding to $R_H$. Additionally, it obtains NPMs from the marginals queried by PrivMRF, and stores them in the initial $\widetilde{\mathcal{M}_{\text{all}}}$ (Line 2). Specifically, for any marginal $M_S$ of $R_H$ such that $H.A_{\text{size}} \in S$. Let $s$ be any possible group size. The cells for $H.A_{\text{size}} = s$ in $M_S$ can be used as an NPM for group size $s$, since both the cells and the NPM are computed from the $R_H$ tuples with $H.A_{\text{size}} = s$, and the normalizing factor $\frac{1}{W_s}$ for the NPM ensures that its total equals $n_s$. For this purpose, we adapt the original PrivMRF algorithm by modifying the step that creates a set of candidate marginals for selection, by removing all candidates that do not include $H.A_{\text{size}}$. This adjustment allows PrivPetal to decompose all

---

**Algorithm 1:** PrivPetal For One Foreign Key Reference

**Input:** Referencing relation $R_I$, referenced relation $R_H$, R-score noise scale $\sigma_R$, NPM noise scale $\sigma_M$, h-score noise scale $\sigma_h$, group size noise scale $\sigma_n$.
**Output:** Synthetic relations $\widetilde{R_H}$ and $\widetilde{R_I}$.

1 Invoke PrivMRF to synthesize $R_H$, producing $\widetilde{R_H}$, which constitutes the columns of $\widetilde{F(R_I, R_H)}$ corresponding to $R_H$;
2 Decompose the noisy marginals queried by PrivMRF into NPMs, and let them be $\widetilde{\mathcal{M}_{\text{all}}}$;
3 Calculate the R-scores $\widetilde{R(\cdot, \cdot)}$ for all attribute pairs, each injected with $\mathcal{N}(0, \sigma_R)$ noise independently // Section 4.2;
4 Count the number of tuples $n_s$ in $F(R_I, R_H)$ for each group size $s \le N$, each injected with $\mathcal{N}(0, \sigma_n)$ noise independently;
5 $\mathcal{A}_{\text{syn}} \leftarrow \mathcal{A}_H$;
6 **for** each attribute $I_i.A_j \in \mathcal{A}_N$ in $\widetilde{F(R_I, R_H)}$ **do**
7    Construct graphical models and update $\widetilde{\mathcal{M}_{\text{all}}}$ by invoking Algorithm 2($I_i.A_j$, $\widetilde{\mathcal{M}_{\text{all}}}$, $\mathcal{A}_{\text{syn}}$, $\widetilde{R(\cdot, \cdot)}$, $\{\widetilde{n_s}\}$, $\sigma_M$, $\sigma_h$);
8    Compute $p_s(I_i.A_j \mid C)$ for each group size $s \ge i$ using the graphical models;
9    Sample $t[I_i.A_j]$ for each tuple $t \in \widetilde{F(R_I, R_H)}$ based on $t[C]$ using $p_{s_t}(I_i.A_j \mid C)$;
10    $\mathcal{A}_{\text{syn}} \leftarrow \mathcal{A}_{\text{syn}} \cup \{I_i.A_j\}$;
11 Decompose $\widetilde{F(R_I, R_H)}$ to obtain $\widetilde{R_I}$;
12 **return** $\widetilde{R_I}, \widetilde{R_H}$;

---

marginals queried by PrivMRF into NPMs, which are subsequently stored in the initial $\widetilde{\mathcal{M}_{\text{all}}}$ (Line 2). [1]

Next, PrivPetal calculates the R-scores [41] (which signify attribute correlations, detailed in Section 4.2), and injects each with Gaussian noise of scale $\sigma_R$ independently to satisfy DP (Line 3). Then, for each possible group size $s \le N$, PrivPetal counts the number of flattened tuples $n_s$ in the FR with group size $s$, and obtains a noisy version $\widetilde{n_s}$ under DP (Line 4). After that, PrivPetal sets the synthesized attributes $\mathcal{A}_{\text{syn}}$ as the household attributes $\mathcal{A}_H$ (Line 5) and sequentially samples individual attributes using graphical models (Lines 6-9). Specifically, for each individual attribute $I_i.A_j \in \mathcal{A}_N$ to be sampled, PrivPetal constructs a graphical model to estimate $p_s(\{I_i.A_j\} \cup C)$ for each group size $s \ge i$ (Line 7), which is elaborated in Section 4.3. Then, for each flattened tuple $t \in \widetilde{F(R_I, R_H)}$, the algorithm uses the corresponding graphical model for group size $s_t = t[H.A_{\text{size}}]$ to calculate a conditional probability $p_{s_t}(I_i.A_j \mid C)$, and samples $t[I_i.A_j]$ based on $t[C]$ (Line 8).

After sampling $I_i.A_j$ for all flattened tuples, PrivPetal adds $I_i.A_j$ to $\mathcal{A}_{\text{syn}}$ (Line 9) and proceeds to sample the next individual attribute. Once all individual attributes are sampled, Algorithm 1 recovers groups from $\widetilde{F(R_I, R_H)}$ to form the synthetic base relation $\widetilde{R_I}$ (Line 10), and returns $\widetilde{R_I}$ and $\widetilde{R_H}$ as the final output.

### 4.2 Selecting Correlated Attributes

Consider the step in PrivPetal, that synthesizes an $I_i.A_j$ in the FR, i.e., the $j$-th attribute of the $i$-th join partner. Let $\mathcal{A}_{\text{syn}}$ be the

---

[1] In fact, initializing $\widetilde{\mathcal{M}_{\text{all}}}$ by reusing the marginals queried by PrivMRF is just one of the various strategies for identifying NPMs capturing correlations among attributes in the FR. We discuss this issue further in Appendix D.

**Algorithm 2:** MRF Construction

**Input:** Target attribute $I_i.A_j$, synthesized attribute set $\mathcal{A}_{\text{syn}}$, noisy NPM set $\widetilde{\mathcal{M}_{\text{all}}}$, noisy R-scores $\widetilde{R(\cdot, \cdot)}$, noisy group sizes $\{\widetilde{n_s}\}$, NPM noise scale $\sigma_M$, h-score noise scale $\sigma_h$.

**Output:** MRF structure $\mathcal{S}$, MRF parameters $\theta$, updated noisy NPM set $\widetilde{\mathcal{M}_{\text{all}}}$.

1 Let $C$ be the set of top $N_{\text{MRF}}$ attributes in $\mathcal{A}_{\text{gen}} \setminus \{H.A_{\text{size}}\}$ with the highest noisy R-scores pertinent to $I_i.A_j$;
2 Let $C_{\text{all}}$ be a set such that (i) each $S \in C_{\text{all}}$ is a subset of $\{I_i.A_j\} \cup C$ and (ii) the NPMs on $S$ are in $\widetilde{\mathcal{M}_{\text{all}}}$;
3 $\mathcal{S} \leftarrow \varnothing$;
4 **for** $t \in [T_1]$ **do**
5  $\quad$ Select $S \in C_{\text{all}}$ with the highest h-score $h(S)$;
6  $\quad$ Insert $S$ into $\mathcal{S}$, and remove $S$ from $C_{\text{all}}$;
7  $\quad$ Estimate the parameters of the MRFs from $\widetilde{\mathcal{M}_{\mathcal{S},s}}$ for each $s \geq i$;
8 Construct a set $C_{\text{new}}$ of candidate attribute sets such that each candidate satisfies $\lambda$-usefulness [41];
9 **for** $t \in [T_2]$ **do**
10 $\quad$ Sample a set $C$ of $k$ attribute sets from $C_{\text{new}}$;
11 $\quad$ Select $S \in C$ with the highest noisy h-score $\widetilde{h(S)}$;
12 $\quad$ Insert $S$ into $\mathcal{S}$, and remove $S$ from $C_{\text{new}}$;
13 $\quad$ $\widetilde{\mathcal{M}_{\text{all}}} \leftarrow$ Algorithm 3($S, \widetilde{\mathcal{M}_{\text{all}}}, \{\widetilde{n_s}\}, \sigma_M$);
14 $\quad$ Estimate the parameters of the MRFs from $\widetilde{\mathcal{M}_{\mathcal{S},s}}$ for each $s \geq i$;
15 **return** MRFs for group sizes $s \geq i$, $\widetilde{\mathcal{M}_{\text{all}}}$;

set of previously synthesized attributes. The algorithm needs to identify a subset $C \subset \mathcal{A}_{\text{syn}}$ of attributes that are correlated to $I_i.A_j$. For any attribute pair $A_1, A_2 \in \mathcal{A}_N$, we use the R-score [41] to measure their correlation, which has been shown to be robust under DP. Specifically, the R-score is the difference between the actual marginal on $\{A_1, A_2\}$ and the marginal assuming independence between $A_1$ and $A_2$. In PrivPetal, these marginals are replaced with NPMs to measure attribute correlations in the permutation relation. Since NPMs are parameterized with the group size $s$, the R-score is adapted to take into account all group sizes:

$$R(A_1, A_2) = \tfrac{1}{2} \sum_s \left\| M_{\{A_1,A_2\},s} - \tfrac{1}{n_s} M_{\{A_1\},s} \otimes M_{\{A_2\},s} \right\|_1, \quad (3)$$

where $M_{\{A_1,A_2\},s}, M_{\{A_1\},s}$, and $M_{\{A_2\},s}$ are the NPMs for group size $s$ on attribute sets $\{A_1, A_2\}, \{A_1\}$, and $\{A_2\}$, respectively. $\|\cdot\|_1$ represent the $L_1$ norm. $\otimes$ represents the outer product, which produces a marginal on $\{A_1, A_2\}$ under the assumption of their independence. The factor $\tfrac{1}{n_s}$ normalizes the product to match the original total.

The above R-score measures the correlation by evaluating the difference between $M_{\{A_1,A_2\},s}$ and $\tfrac{1}{n_s} M_{\{A_1\},s} \otimes M_{\{A_2\},s}$. The summation over $s$ ensures the generality across different group sizes. Based on the R-scores, PrivPetal selects the correlated attribute set $C$ as the top-$N_{\text{MRF}}$ ($N_{\text{MRF}}$ is system parameter) attributes in $\mathcal{A}_{\text{syn}} \setminus \{H.A_{\text{size}}\}$ with the highest noisy R-scores pertinent to $I_i.A_j$.

### 4.3 Building Graphical Models

**Building Markov random fields (MRFs).** Following the state-of-the-art single-relation synthesis method PrivMRF [6], PrivPetal uses MRF as its graphical model, for computing each $p_s(C \cup \{I_i.A_j\})$ from a given set of NPMs $\mathcal{M}_{\mathcal{S},s} = \{M_{S,s} \mid S \in \mathcal{S}\}$, where $\mathcal{S}$ is the set of attribute sets corresponding to these NPMs (the selection of these NPMs is detailed soon). For any possible tuple $t \in \mathcal{X}_{C \cup \{I_i.A_j\}}$, the MRF defines the distribution value $p_s(t)$ as $p_s(t) \propto \prod_{S \in \mathcal{S}} \exp(\theta_S(t[S]))$, where $\theta_S$ is a vector of parameters associated with the attribute set $S$, and $\theta_S(t[S])$ represents the parameter allocated for $t[S]$. The distribution values are normalized to ensure that $\sum_t p_s(t) = 1$. We refer to $\mathcal{S}$ as the *structure* of the MRF (which is the attribute sets of the given NPMs $\mathcal{M}_{\mathcal{S},s}$ in PrivPetal) and denote the concatenated parameter vector by $\theta = (\theta_S)_{S \in \mathcal{S}}$, which is estimated from $\mathcal{M}_{\mathcal{S},s}$ through maximum likelihood estimation [22].

Next, we detail how PrivPetal constructs an MRF for each individual attribute $I_i.A_j$ and each group size $s = i, \ldots, N$, to produce $p_s(C \cup \{I_i.A_j\})$. It suffices to consider the case that the group size $s \geq i$, since otherwise (i.e., $s < i$) $I_i.A_j$ is always NULL. We follow the general framework of PrivMRF [6], adapting it to handle NPMs. Algorithm 2 presents the pseudocode for MRF construction in PrivPetal, which comprises three steps. First, it selects a set of correlated attributes $C$ (Line 1), as described in the previous subsection. Second, it runs $T_1$ iterations to select queried NPMs in $\widetilde{\mathcal{M}_{\text{all}}}$ (represented by the set $C_{\text{all}}$ of corresponding attribute sets) to construct the MRFs (Lines 2-7). Third, it runs $T_2$ iterations to select new permutation marginals yet to be queried to further refine the MRFs (Lines 8-14). Finally, the algorithm returns the MRFs for all group sizes $s \geq i$ and the updated noisy NPM set $\widetilde{\mathcal{M}_{\text{all}}}$ (Line 15).

The second and third steps follow Algorithm 4 from PrivMRF [6], which iteratively finds the marginal with the largest error, and add it to the MRF to mitigate the error. Specifically, for any candidate attribute set $S$, the error is measured using the h-score [6]:

$$h(S) = \sum_{s: s \geq i} \| M_{S,s} - p_{S,s} \|_1,$$

where $p_{S,s}$ is the marginal distribution of $p_s$ on $S$, and is calculated from the MRF through standard algorithms (e.g., junction tree algorithm). To preserve DP, these h-scores are injected with Gaussian noise of scale $\sigma_h$ to preserve DP, resulting in the noisy h-scores $\widetilde{h(S)}$. Once an attribute set $S$ is added to the structure $\mathcal{S}$, Algorithm 2 updates the parameters of all MRFs using the corresponding permutation marginals $\widetilde{\mathcal{M}_{\mathcal{S},s}} \subset \widetilde{\mathcal{M}_{\text{all}}}$ (Lines 7 and 14). The hyperparameters $T_1, \lambda, k$ from PrivMRF are set to be their default values, except for $T_2$, which we set to 1 because the Step 2 already yields high-quality MRFs, and only a single iteration is needed to select an additional permutation marginal.

Compared to PrivMRF, Algorithm 2 introduces three non-trivial adaptations. First, the computation of the correlated attributes $C$ is based on the method explained in Section 4.2. Second, it employs a novel module for querying NPMs in Line 13, elaborated soon. Third, it uses the noisy permutation marginals $\widetilde{M_{S,s}}$ instead of the true marginals $M_{S,s}$ when computing the h-score in Line 5. This avoids direct access to private data, thereby reducing the privacy cost.

**Querying NPMs.** Algorithm 3 presents the pseudocode for the proposed module for querying NPMs. Among its inputs, $S_T$ is a subset of the attribute set $\mathcal{A}_N$ of $F(R_I, R_H)$, and contains at most $o$ different individual identifiers. The set $\widetilde{\mathcal{M}_{\text{all}}}$ is used to store queried NPMs. The algorithm aims to query $M_{S_T,s}$ for each possible group

**Algorithm 3:** NPM Querying

**Input:** Target attribute set $S_T$, NPM set $\widetilde{\mathcal{M}_{\text{all}}}$, noisy group sizes $\{\widetilde{n_s}\}$, noise scale $\sigma$.
**Output:** Updated NPM set $\widetilde{\mathcal{M}_{\text{all}}}$.

1. Let $D$ be the number of different individual identifiers in $S_T$;
2. Identify $S$ satisfying $S \to S_T$ in permutation relations;
3. **for** each group size $s = D, \dots, N$ **do**
4. 　　Count $M_{S,s}$ from the permutation relation $P_s(R_I, R_H)$;
5. **if** the group size merging interval $I$ is non-empty **then**
6. 　　Merge $M_{S,s}$ over $I$ and inject independent noise $\mathcal{N}(0, \sigma)$ into each merged count;
7. 　　Rescale the noisy counts to the original total to obtain $\widetilde{M_{S,s}}$ for each $s \in I$;
8. Inject independent noise $\mathcal{N}(0, \sigma)$ into each count of $M_{S,s}$ to obtain $\widetilde{M_{S,s}}$ for each $s \in \{D, \dots, N\} \setminus I$;
9. **for** each group size $s = D, \dots, N$ **do**
10. 　　**for** each $S' \subset \mathcal{A}_s$ satisfying $S \to S'$ **do**
11. 　　　　$\widetilde{M_{S',s}} \leftarrow \widetilde{M_{S,s}}$;
12. 　　　　Roll up $\widetilde{M_{S',s}}$ to smaller NPMs;
13. 　　　　Add all the noisy NPMs to $\widetilde{\mathcal{M}_{\text{all}}}$;
14. **return** $\widetilde{\mathcal{M}_{\text{all}}}$;

**Algorithm 4:** PrivPetal for Multiple Foreign Keys

**Input:** Database $\mathcal{R}$.
**Output:** Synthetic database $\widetilde{\mathcal{R}}$.

1. **for** each public relation $R \in \mathcal{R}$ **do**
2. 　　$\widetilde{R} \leftarrow R$;
3. Create a list $L$ of private relations in $\mathcal{R}$ by applying a topological sort to the graph of private foreign key references and reverse $L$;
4. **for** each relation $R \in L$ **do**
5. 　　**if** $R$ does not have any foreign keys **then**
6. 　　　　Invoke PrivMRF to synthesize $R$, producing $\widetilde{R}$;
7. 　　**else**
8. 　　　　**for** each foreign key $\text{FK}(R, R')$ of $R$ **do**
9. 　　　　　　Invoke Algorithm 1 with $R$ and $R'$, producing $\widetilde{R}$;
10. 　　　　Merge all resulting $\widetilde{R}$ by merging their foreign keys;
11. Set $\widetilde{\mathcal{R}}$ as the collection of all $\widetilde{R}$;
12. **return** $\widetilde{\mathcal{R}}$;

size $s$, and find all other NPMs that can be represented by $M_{S_T,s}$. It stores all these NPMs in $\widetilde{\mathcal{M}_{\text{all}}}$ for later use.

Let $D$ be the number of different individual identifiers in $S_T$ (Line 1). Algorithm 3 only queries $M_{S_T,s}$ for $s \geq D$. This is because $M_{S_T,s}$ contains more individuals than flattened tuples of group size $s < D$, and cannot be used to synthesize them. Algorithm 3 identifies the attribute set $S$ satisfying $S \to S_T$ in the permutation relation $P_s(R_I, R_H)$ (Line 2). Then, for each group size $s = D, \dots, N$, it counts the NPM $M_{S,s}$ from $P_s(R_I, R_H)$ (Line 4).

Algorithm 3 then injects noise into $M_{S,s}$ for each $s$ to preserve DP. Note that the total of $M_{S,s}$ equals $n_s$. If $n_s$ is sufficiently large, Algorithm 3 directly injects Gaussian noise of scale $\sigma$ into the counts of $M_{S,s}$ (Line 8). Otherwise, a group size merging interval $I$ should be given, and Algorithm 3 merges NPMs across $I$ to improve their resilience to the DP noise (Lines 5-6). It then rescales the merged NPM to the original totals (Line 7). Specifically, for each attribute value combination $x \in \mathcal{X}_S$ and each group size $s \in I$, we have:

$$\text{MergedCount} = \sum_{s' \in I} M_{S,s'}[x],$$

$$\text{NoisyCount} = \text{MergedCount} + \mathcal{N}(0, \sigma),$$

$$\widetilde{M_{S,s}}[x] = \frac{\widetilde{n_s}}{\sum_{s' \in I} \widetilde{n_{s'}}} \text{NoisyCount},$$

where $\widetilde{n_s}$ is the noisy version of $n_s$, as required by DP. Rescaling with $\widetilde{n_s}$ ensures that merged NPMs maintain the original totals. Intuitively, Algorithm 3 merges the counts over group sizes, and for each $x \in \mathcal{X}_S$ separately. Merged NPMs capture the general correlations for different group sizes and, while potentially introducing some bias, are less susceptible to noise.

Regarding the choice of $I$, since larger group sizes are often less frequent in real-world data, an optimization is to merge all sizes exceeding a threshold. Alternatively, we can determine the optimal interval by comparing the expected errors of the merged NPMs for different intervals. For brevity, we leave this for future work, and explain the merging in our experiments in Section 6.

After noise injection, Algorithm 3 enumerates all attribute sets $S' \subset \mathcal{A}_s$ satisfying $S \to S'$ for each $s = D, \dots, N$. It then copies $\widetilde{M_{S,s}}$ to $\widetilde{M_{S',s}}$, projects them into smaller NPMs, and stores all these NPMs in $\widetilde{\mathcal{M}_{\text{all}}}$ (Lines 9-13). We note that the enumeration of $S'$ includes $S_T$, as we also have $S \to S_T$. Finally, the updated set $\widetilde{\mathcal{M}_{\text{all}}}$ contains all these noisy NPMs and is returned for later use (Line 14).

### 4.4 Handling Multiple Foreign Keys

So far, our discussions have focused on the scenario with two relations $R_I$ and $R_H$ and one foreign key $\text{FK}(R_I, R_H)$. Next, we extend PrivPetal to the general case with multiple relations, including one primary private relation and several other secondary private relations depending on it. Given such a database $\mathcal{R}$, Algorithm 4 presents the pseudocode of PrivPetal for generating a synthetic version $\widetilde{\mathcal{R}}$. Since releasing public relations $R$ does not incur any privacy cost, they are denoted as $\widetilde{R}$ and are preserved in the synthetic database (Lines 1-2). The algorithm then focuses on synthesizing private relations and their foreign keys.

The main idea is to synthesize private relations sequentially, and apply Algorithm 1 to process each of the foreign keys. Since the private foreign key references form a DAG, Algorithm 4 applies a topological sort on the private relations to produce a list $L$ of relations, and reverses $L$ (Line 3). The algorithm then synthesizes private relations in the order of $L$, by which a referenced relation (e.g., $R_H$) always precedes its referencing relations (e.g., $R_I$).

For each private relation $R$ in $L$, if $R$ does not have any foreign key, PrivPetal simply invokes PrivMRF to generate its synthetic version $\widetilde{R}$ (Line 6). Otherwise, PrivPetal processes the foreign keys sequentially: for each such $\text{FK}(R, R')$, PrivPetal invokes Algorithm 1 with $R$ as $R_I$ and $R'$ as $R_H$ (Line 9). As $R$ references $R'$, the synthetic version $\widetilde{R'}$ is either synthesized in previous iterations according to the order of $L$ or is a public relation released directly. Thus, we modify Algorithm 1 to avoid regenerating $\widetilde{R'}$ by removing Lines 1-2, and replacing the synthesized $\widetilde{R_H}$ with the existing $\widetilde{R'}$.

One remaining issue is that we need an alternative method to obtain the initial NPM set $\widetilde{\mathcal{M}_{\text{all}}}$ after removing Lines 1-2 in Algorithm 1. Specifically, if $\widetilde{R'}$ is synthesized by the previous iterations of Algorithm 4 using PrivMRF, we decompose its marginals into NPMs and store them in the initial $\widetilde{\mathcal{M}_{\text{all}}}$. Otherwise (i.e., $\widetilde{R'}$ is released as a public relation), we apply PrivMRF to $R'$ to obtain the NPMs, without accounting for any privacy cost, as $R'$ does not contain any private information.

When relation $R$ has multiple foreign keys $FK(R, R')$, multiple $\widetilde{R}$ are synthesized during the above process, with each $\widetilde{R}$ having its respective $FK(R, R')$. To reconcile the differences among these $\widetilde{R}$ and merge their foreign keys, we synthesize the first $\widetilde{R}$ without any modifications and adjust the subsequent invocations of Algorithm 1. Specifically, in Line 8 of Algorithm 1, only tuples in the first $\widetilde{R}$ are allowed to be sampled, ensuring that all $\widetilde{R}$ share the same collection of tuples. Algorithm 4 then merges all $\widetilde{R}$ and their foreign keys into a single $\widetilde{R}$ by matching their tuples (Line 10). Finally, the algorithm finishes by returning the synthetic database $\widetilde{\mathcal{R}}$, which is the collection of all synthetic relations (Lines 11-12).

**Discussion.** PrivPetal strives to generate a synthetic database that statistically mimics the input database, and preserves various dependencies and integrity constraints. For row-level dependencies, when rows in a relation follow different distributions conditioned on an attribute, such as YEAR, PrivPetal can capture inter-attribute correlations specific to the YEAR attribute. It then synthesizes different tuples for each year, preserving the corresponding row-level dependencies. Similarly, for functional dependencies $X \to Y$, PrivPetal seeks to preserve these by capturing the inter-attribute correlations for $X$ and $Y$. The same case applies to multivalued dependencies, provided PrivPetal captures all inter-attribute correlations related to these dependencies. We note that, however, the preservation of such dependencies is not guaranteed, since PrivPetal may omit inter-attribute correlations that are either weak or involve too many attributes. Furthermore, the synthetic database may deviate from these dependencies due to the noise introduced by DP.

Regarding integrity constraints, PrivPetal enforces primary key constraints by assigning a unique primary key value to each tuple and enforces foreign key constraints by generating values that correspond exclusively to existing primary keys. Additionally, NOT NULL and CHECK constraints are enforced by limiting the domains of attributes. One limitation of PrivPetal is that it cannot effectively enforce composite key constraints, since the method cannot ensure the uniqueness of each combination of attribute values in a composite key.

## 5 Privacy Analysis

In this section, we demonstrate that PrivPetal satisfies $(\epsilon, \delta)$-DP. We begin by introducing the concept of $L_2$ sensitivity and a foundational result on the application of Gaussian noise.

*Definition 5.1 ($L_2$ Sensitivity [11]).* Let $f$ be a function that maps a database to a real vector. The $L_2$ sensitivity of $f$, denoted as $\Delta(f)$, is the maximum value of $\|f(\mathcal{R}) - f(\mathcal{R}')\|_2$ for any two neighbor databases $\mathcal{R}$ and $\mathcal{R}'$, where $\|\cdot\|_2$ denotes the $L_2$ norm.

Theorem 5.2 (Analytic Gaussian Mechanism [3]). *Let $\{f_1, \ldots, f_k\}$ be a set of functions. For any $i = 1, \ldots, k$, suppose that we inject independent Gaussian noise $\mathcal{N}(0, \sigma_i^2)$ into each element in $f_i$'s output. Then, the perturbed functions as a whole satisfy $(\epsilon, \delta)$-DP, iff*

$$\Phi\left(\frac{\gamma}{2} - \frac{\epsilon}{\gamma}\right) - e^\epsilon \cdot \Phi\left(-\frac{\gamma}{2} - \frac{\epsilon}{\gamma}\right) \leq \delta, \tag{4}$$

*where $\Phi$ is the cumulative distribution function of the standard normal distribution, and*

$$\gamma = \sqrt{\sum_{i=1}^{k} \left(\frac{\Delta(f_i)}{\sigma_i}\right)^2}. \tag{5}$$

Since PrivPetal injects Gaussian noise into all its query results, we leverage this theorem to demonstrate its DP guarantee. Given the desired privacy parameters $\epsilon$ and $\delta$, we find the largest $\gamma$ such that Eq. (4) is satisfied, and then select the noise scales $\sigma_i$ according to Eq. (5). Specifically, we refer to $\sum_{i=1}^{k} \left(\frac{\Delta(f_i)}{\sigma_i}\right)^2$ as the *privacy cost* associated with $\{f_i\}$. We then quantify the privacy costs of the algorithms using noise scales, and ensure that the square root of the total privacy costs is smaller than $\gamma$.

Consider using Algorithm 4 to synthesize a database. Neighbor databases differ by adding or removing a tuple from the primary private relation and all its dependent tuples, as described in Section 2.1. The number of added or removed tuples can vary for different foreign keys $FK(R, R')$, leading to different privacy costs for the subroutines of PrivPetal. To quantify these costs, we let $\tau$ denote the maximum number of tuples in $R'$ that can change.

Next, we quantify the privacy cost of each algorithm. Both Algorithm 1 and Algorithm 4 use PrivMRF as a subroutine. As PrivMRF also uses the analytic Gaussian mechanism, we can denote its privacy cost by $C_H$, as given by Lemma 2 in [6]. We then have the following theorems, with proofs deferred to Appendix B and Appendix C.

Theorem 5.3. *The privacy cost of applying PrivMRF to $R'$ is:*

$$C_{\text{PrivMRF}} = \tau^2 C_H. \tag{6}$$

Theorem 5.4. *The privacy cost of applying Algorithm 1 to $R$ and $R'$, with Lines 1-2 removed, is:*

$$C_{\text{one-FK}} = 2\tau^2 \frac{|\mathcal{A}_H|^2 + 2|\mathcal{A}_H||\mathcal{A}_I| + 2|\mathcal{A}_I|^2 - |\mathcal{A}_H|}{\sigma_R^2}$$

$$+ \frac{\tau^2}{\sigma_n^2} + \tau^2 T_2 N |\mathcal{A}_I| \left(\frac{k}{\sigma_h^2} + \frac{1}{\sigma_M^2}\right). \tag{7}$$

Finally, the total privacy cost of Algorithm 4 is given by summing the privacy costs of applying PrivMRF and applying Algorithm 1. For example, in the simplest case where the database contains only two private relations $R_I$ and $R_H$, the total privacy cost is $C_{\text{PrivMRF}} + C_{\text{one-FK}}$, with $\tau = 1$.

For the distribution of the privacy cost in Algorithm 1, we allocate the three summation terms in Eq. (7) in a 1 : 1 : 8 ratio. This is because the third term is used for constructing MRFs with Algorithm 2, and should incur the highest privacy cost. Additionally, we let $\frac{k}{\sigma_h^2} : \frac{1}{\sigma_M^2} = 1 : 9$ because h-scores are used for selecting NPMs, while NPMs are used for capturing correlations directly and are more informative. In Algorithm 4, the distribution of privacy costs

Table 5: Statistics of census datasets.

(a) California

| Relation | # records | # attributes | domain size |
|---|---|---|---|
| Individual | 1,690,642 | 23 | $\approx 6.77 \times 10^{12}$ |
| Household | 616,115 | 10 | $\approx 3.24 \times 10^{7}$ |

(b) Île-de-France

| Relation | # records | # attributes | domain size |
|---|---|---|---|
| Individual | 4,297,133 | 14 | $\approx 1.84 \times 10^{10}$ |
| Household | 1,911,412 | 10 | $\approx 1.24 \times 10^{7}$ |

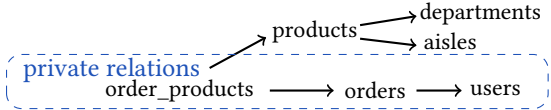

Figure 4: Foreign key dependencies in Instacart.

among the invocations of PrivMRF and Algorithm 1 depends on the scales of relations and the overall database schema. In particular, the number of NPMs queried by PrivPetal increases linearly with the number of attributes in the relations. Consequently, the scale of noise also increases, resulting in a higher error in the synthetic database. To mitigate this problem, relations with more attributes should be allocated a higher privacy cost. Furthermore, when the input database contains a larger number of foreign keys, the privacy cost allocated to each relation decreases, which can lead to increased error. Particularly, relations located downstream in the DAG of the foreign key dependencies should be allocated a higher privacy cost, as queries on these relations usually incur higher sensitivity. We explain the specifics of privacy budget allocation in our experiments in Section 6.4.

## 6 Experiments

We evaluate the performance of PrivPetal using census datasets, the Instacart dataset, and the TPC-H benchmark, on a Linux server configured with a 32-core 3.0GHz CPU. Each run of PrivPetal was completed within 11 hours. Section 6.1 describes the experimental setup. Sections 6.2, 6.3 and 6.4 present the evaluation results.

### 6.1 Setup

**Datasets.** We use two real-world census datasets collected from [8, 34]. Their statistics are presented in Table 5. Both the California and Île-de-France datasets contain an individual relation and a household relation, with each individual referencing a household. We set the household relation as the primary private relation, and the individual relation as the secondary private relation.

The Instacart dataset [35] is a real-world e-commerce dataset from a grocery ordering and delivery application, and comprises approximately 3,000,000 orders and 200,000 users. It includes six tables with five foreign key dependencies, as shown in Figure 4. We assume that the users relation is the primary private relation, and orders and order_products are secondary private relations, resulting in three private foreign keys in total.

For the TPC-H benchmark [1], its schema is depicted in Figure 5. Orders and Lineitem are designated as the primary private relation and the secondary private relation, respectively, resulting in three private foreign keys. The original TPC-H benchmark generates tuples and foreign keys using uniform distributions. Therefore, the correlations in the database are negligible, making it unsuitable for evaluating synthesis methods. To address this problem, we adopt similar modifications as in [7] to introduce significant correlations into TPC-H. This ensures a fair comparison between PrivPetal and PrivLava [7], and highlights their capability to synthesize correlated relational data. Specifically, we apply the following modifications:

(1) We assign a weight to the P_TYPE attribute in Part, ranging from [0.2, 1], and another weight to the P_BRAND attribute, ranging from [0.2, 2]. The L_QUANTITY attribute in Lineitem is then multiplied by the product of these weights. This step introduces inter-relational correlations between Lineitem and PartSupp, as well as intra-group correlations among lineitems that share the same PartSupp.
(2) For each year, we redistribute the orders such that 40% of them refer to 15% of the customers, while the remaining 60% of orders refer to the other 85% of customers. This creates an inter-relational correlation between Customer and Orders.
(3) For each year $y$, we drop orders with a 75% probability, except for those containing $(y - 1992)$ lineitems. This introduces a correlation between order size and order year.
(4) If the O_ORDERPRIORITY attribute in Orders is 1-URGENT or 2-HIGH, then the L_SHIPMODE attribute in Lineitem has a 50% chance of being AIR or REG AIR. Otherwise, that probability is reduced to 10%. This induces an inter-relational correlation between Orders and Lineitem.

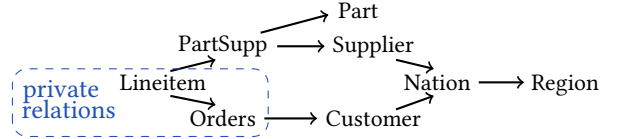

Figure 5: Foreign key dependencies in TPC-H.

**Baselines.** PrivLava [7] is the state-of-the-art method for synthesizing multi-relation databases under DP. We adopt PrivLava alongside three DP single-relation synthesis methods as baselines: PrivMRF [6], PB-PGM [30], and PrivBayes [41]. In addition, we modify these single-relation methods to synthesize foreign keys. First, we use 80% of the privacy cost to create synthetic versions for private relations. Then, we generate a noisy histogram $\widetilde{H}$ recording the group size distribution for each private foreign key FK($R, R'$), with the remaining 20% of the privacy cost. Finally, we randomly link the synthetic tuples in $R$ with those in $R'$, ensuring that the group size distribution in $\widetilde{H}$ is maintained.

**Privacy Parameters.** We vary the privacy parameter $\epsilon$ for $(\epsilon, \delta)$-DP, fixing $\delta$ to the inverse of the number of tuples in the secondary private relation, which aligns with previous work [4, 6, 7, 13, 19].

### 6.2 Results on California

In this experiment, we evaluate the performance of each method based on the utility of the synthetic census datasets. Our evaluation, aligned with the approach in PrivLava [7], involves a set of queries on both the synthetic and original data. The utility of the synthetic data is then measured by the relative errors of these queries. Evaluations on more complex tasks, such as machine learning or exploratory data analytics, are left as future work.

Each query, denoted by $Q(s, P_H, c, \{P_i\})$, counts the number of households $h$ in data, based on three criteria: (i) the household contains a total of $s$ individuals; (ii) the household meets a predicate $P_H$ on its attributes; and (iii) the household includes $c$ individuals $p_1, \ldots, p_c$, such that each $p_i$ meets a predicate $P_i$ on its attributes.

We use queries that have $c = 1$ and $c = 2$, with a randomly selected $s$. For $P_H$ and $\{P_i\}$, we use 1- and 2-attribute conjunctive predicates. Specifically, we randomly select each household/individual attribute $A$, and a subset $S_A$ from its domain $\mathcal{X}_A$, avoiding duplicated attributes in a predicate. Then, a tuple $t$ meets the predicate if $t[A] \in S_A$ for each $A$. We set $S_A$ such that the overall selectivity of attributes is 0.2 when the query is imposed on uniformly distributed data. This is achieved by setting each $|S_A| = \lfloor (0.2)^{\frac{1}{k}} |\mathcal{X}_A| \rfloor$, where $k$ is the total number of attributes in the query.

Let $\mathcal{R}$ be the input census database, and $\widetilde{\mathcal{R}}$ be a synthetic census database generated by a DP synthesis method. The error of $Q$ indicates if $\widetilde{\mathcal{R}}$ preserves the inter-relational correlations of $\widetilde{\mathcal{R}}$ as $Q$ involves both household attributes and individual attributes. Additionally, it indicates if $\widetilde{\mathcal{R}}$ preserves the intra-group correlations when $c > 1$, and the inter-attribute correlations when the predicates are 2-attribute conjunctive predicates. We evaluate the performance of the synthesis methods by the relative error of $\widetilde{\mathcal{R}}$ w.r.t $Q$:

$$\frac{\text{absolute error of } Q}{\max\{\text{actual result of } Q, 0.01n\}}, \quad (8)$$

where $n$ is the number of households in $\mathcal{R}$, and $0.01n$ is a regularization term to mitigate the impact of very small query results.

For PrivPetal, we adopt order-3 permutation relations and set the group size merging interval $I$ as $[5, +\infty)$. The relative errors for the California and Île-de-France datasets are presented in Figures 6 and 8, respectively. The results are averaged over ten runs, each with 10,000 queries. To accommodate the wide range of errors produced by different methods, we break the y-axis and omit intermediate ticks. These results demonstrate that PrivPetal significantly outperforms all baselines, often by a considerable margin. The errors for single-relation methods remain constant regardless of the variations of $\epsilon$, indicating their inability to capture correlations induced by foreign keys. Conversely, the errors for PrivPetal decrease as $\epsilon$ increases, demonstrating its effective modeling for foreign keys. Furthermore, when $\epsilon$ is sufficiently large (e.g., $\epsilon = 1.60, 3.20$), the errors of PrivPetal are much smaller than those of PrivLava, indicating a more nuanced modeling for foreign keys.

In addition, to clearly demonstrate the correlations captured by the synthesis methods, we measure the Pearson correlation coefficients for continuous attributes in the synthetic data and present these results in Figure 7. The black dashed lines indicate the coefficients measured on the original input data. The closer the other methods' correlation coefficients are to these black dashed lines, the better the synthesis methods, as it shows that correlations in the synthetic data retain the same strength as in the original data. Specifically, for inter-relational correlations, we join the synthetic individual relation and household relation, and measure the Pearson correlation coefficients of two continuous attributes taken respectively from these two relations, as presented in Figures 7 (a)(b)(c). For intra-group correlations, we apply a self-join to the synthetic individual relation, and measure the Pearson correlation coefficients of one continuous attribute from the two joined parts, as shown in Figure 7 (d). This presents the correlations of attributes from co-existing individuals. These results consistently indicate that PrivPetal's correlation coefficients lie closest to the true correlation coefficients, thereby outperforming all baselines. This aligns with the findings from our earlier evaluation based on relative errors of queries.

### 6.3 Results on Instacart

In this experiment, we evaluate the performance of all methods on the Instacart dataset by applying the same relative error metric defined in Eq. (8), where referencing tuples are treated as individual tuples and referenced tuples are treated as household tuples. Compared with the census datasets, the Instacart dataset contains significantly larger groups of referencing tuples. As such, the random predicates $Q$ are much easier to be satisfied, resulting in uniformly high selectivity for all methods. Consequently, to better distinguish the quality of the generated synthetic data, we reduce the overall selectivity of attributes from 0.2 to 0.01.

The results are presented in Figures 9 and 10. Specifically, Figure 9 illustrates the relative errors of all methods measured from the `order_products` and `orders` relations, while Figure 10 shows the relative errors measured from the `orders` and `users` relations. We observe that PrivPetal consistently outperforms all other methods by a substantial margin, further demonstrating its superiority in generating high-utility synthetic data under DP.

### 6.4 Results on TPC-H

In this experiment, we use 8 aggregate queries: Q4, Q5, Q7, Q9, Q12, Q14, Q17, and Q19 from the TPC-H benchmark to assess the performance of each method with relative errors. The regularization term is excluded, as the query answers are all of significant magnitude. For PrivPetal, we use 50% of the privacy cost for the foreign key from `Orders` to `Customer` since the number of tuples in `Customer` is small. Then, each of the remaining two foreign keys shares 25% of the privacy cost. This allocation avoids hyperparameter tuning that may disclose private information about the data.

We adopt 3-order permutation relations and set the group size merging interval $I$ as $[2, +\infty)$ for PrivPetal. The relative errors of all methods averaged over ten runs, are presented in Figure 11. Consistent with the experimental results of the census datasets, PrivPetal outperforms its competitors significantly by a large margin in most cases. This demonstrates its superiority in synthesizing databases with foreign keys under DP. PrivLava performs poorly on Q7, Q14, Q17, and Q19 when $\epsilon = 0.2, 0.4$. This observation supports our claim that PrivLava struggles to maintain data utility when the privacy budget is highly restricted. Meanwhile, the errors of PrivPetal and PrivLava regarding Q4 are close. This is because Q4 mainly depends on the accuracy of group sizes, which are effectively captured by both PrivPetal and PrivLava. In addition, PrivPetal and PrivLava exhibit comparable errors regarding Q9 when $\epsilon = 1.6, 3.20$. The reason is that Q9 depends on the two foreign keys of `Lineitem`. Both PrivPetal and PrivLava rely on merging synthetic relations to merge foreign keys, which equally affects their capability to precisely preserve correlations.

Since TPC-H does not contain continuous attributes across any two relations linked by a foreign key, we cannot measure inter-relational correlations using Pearson correlation coefficients. Instead, we focus on intra-group correlations, as shown in Figure 12. Following the approach in Section 6.2, we perform a self-join of the `Lineitem` relation on its `PART_KEY`, then measure the correlation among L_QUANTITY values that share the same PART_KEY. The results demonstrate that PrivPetal outperforms all baselines with a large margin, as its coefficients are the closest to the true values in all cases.

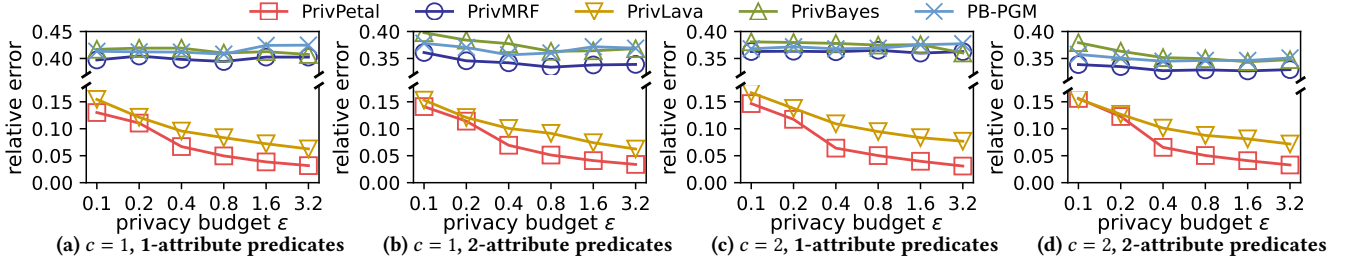

Figure 6: California: relative error vs. $\epsilon$.

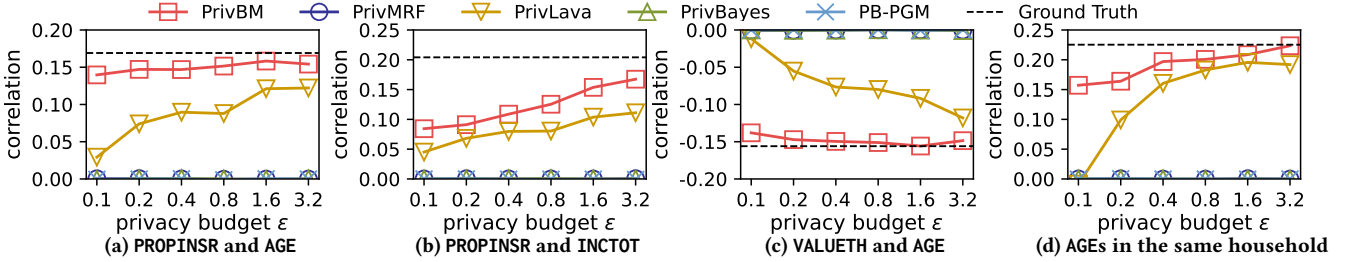

Figure 7: California: Pearson correlation coefficient vs. $\epsilon$.

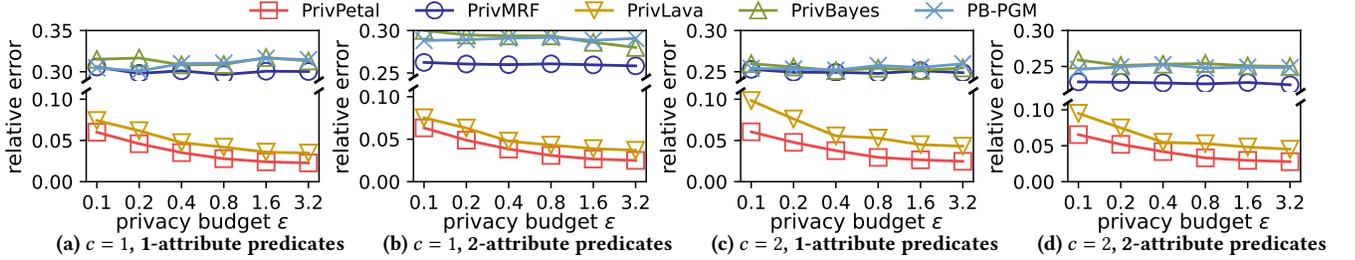

Figure 8: Île-de-France: relative error vs. $\epsilon$.

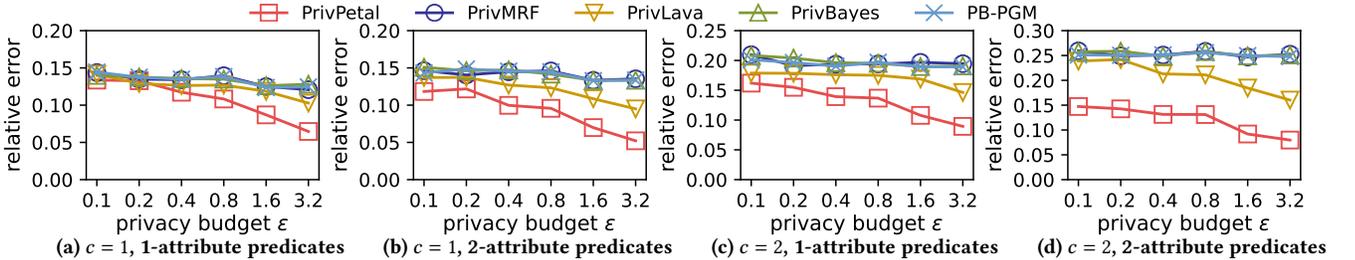

Figure 9: Insacart: relative error of `order_products` and `orders` tables vs. $\epsilon$.

## 7 Related Work

There has been significant progress in synthesizing single-relation databases under DP. Among these, marginal-based methods [5, 6, 9, 25, 29, 30, 33, 41–43] have achieved state-of-the-art results. These methods employ statistical models constructed from marginals to generate synthetic data under DP. For example, PrivBayes [41] constructs a Bayesian network using marginal distributions, while PGM [30] uses MRFs to mitigate the inconsistencies in marginals caused by the noise of DP. More recently, PrivMRF [6] introduces a structure-learning framework for MRFs under DP. However, these methods are limited to single-relation databases and do not account for the complex inter-relational and intra-group correlations introduced by foreign keys.

There are also non-marginal-based methods [17, 18, 20, 24, 27, 31, 32, 37, 37–40, 44] for synthesizing single-relation databases. For instance, MWEM [17] assumes a workload of queries as input and improves the quality of the synthetic data by optimizing the errors of the workload. Alternatively, Table-GAN [32] utilizes generative adversarial networks to synthesize relations. Although these methods provide valuable contributions to the toolkit for DP synthesis, they similarly fall short in handling foreign keys.

Regarding synthesizing databases with foreign keys under DP, PrivLava [7] proposes to group tuples based on foreign key values and classifies them into different types. Then, it synthesizes relations by synthesizing groups corresponding to each type. However, PrivLava incurs a high privacy cost as it needs to refine the types accurately in an iterative manner. Additionally, it must define a large number of types to represent a variety of groups, which increases the difficulty of learning each type. Our method, on the contrary, synthesizes relations linked by foreign keys with possible permutations of tuples. This strategy avoids the high privacy cost for learning types and provides a more granular method for synthesizing relations.

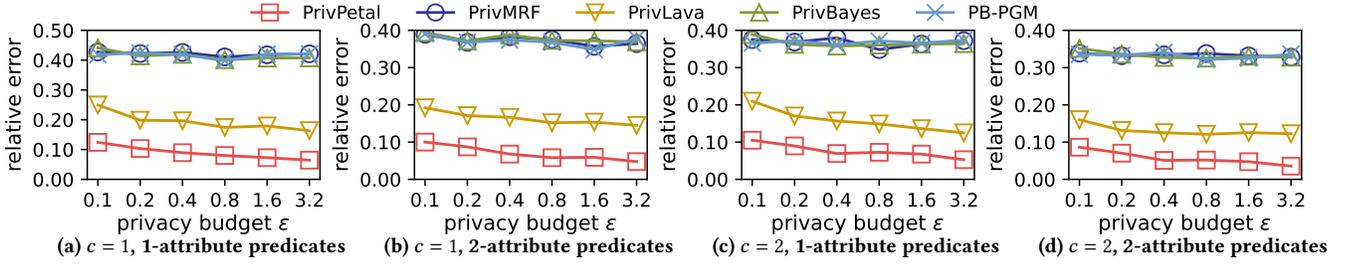

Figure 10: Insacart: relative error of orders and users tables vs. $\epsilon$.

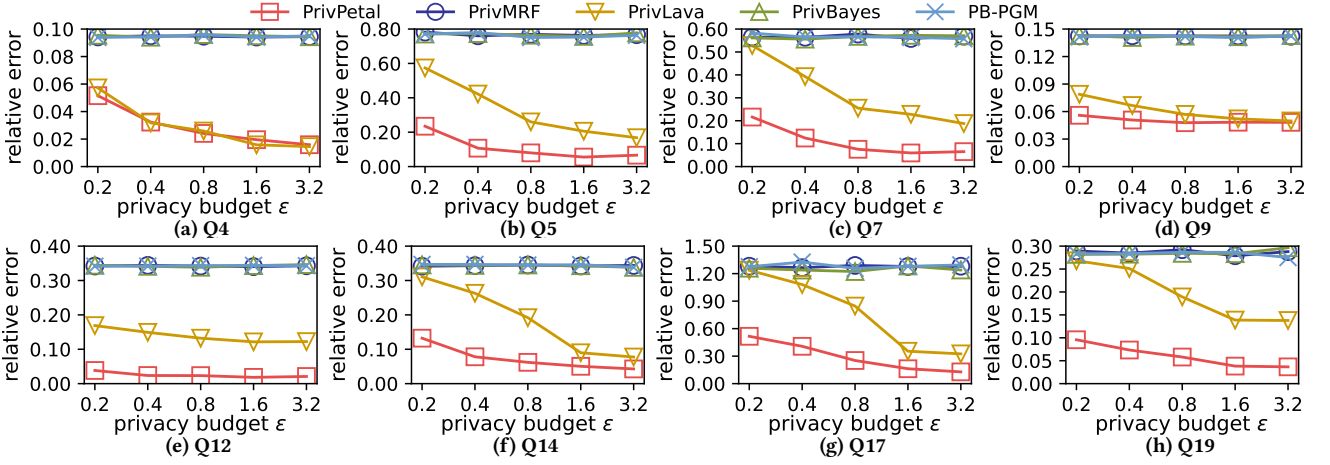

Figure 11: TPC-H: relative error vs. $\epsilon$.

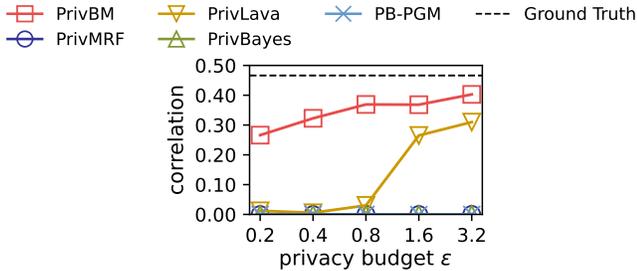

(a) `L_QUANTITY`s in the same `PART`

Figure 12: TPC-H: Pearson correlation coefficient vs. $\epsilon$.

Notably, Ghazi et al. [14] propose a novel algorithm for releasing relational data and establishing lower bounds on the errors of their queries, demonstrating that their algorithms achieve near-optimal performance. Although this work significantly advances the field of DP data synthesis, it is not directly comparable with PrivPetal due to fundamental differences in its DP definition and ours. Specifically, the privacy notion in [14] defines databases as neighboring if one can be converted into the other by adding or removing a single tuple, without considering foreign key constraints. In contrast, our definition assumes that deleting one tuple has the cascading effect of removing all tuples dependent on it. Both definitions have been used in the literature of DP, and our privacy notion provides stronger privacy protection, as we demonstrate in Appendix E with analysis and experiments. Additionally, [14] is a theoretical study focusing on synthesizing data to answer linear queries over multi-table joins, whereas PrivPetal is task-agnostic and applicable to a broad range of applications.

There are also non-DP synthesis methods for relational databases with foreign keys. Previous work [13, 15] uses predefined constraints to guide the generation of foreign keys. However, the derivation of these constraints under DP remains unresolved. One approach [26] proposes to use a parametric factor graph to model the database and sample tuples from the model to obtain the synthetic data. Other approaches [21, 23, 28] preserve the original foreign keys while perturbing other attributes to protect privacy. These methods, however, do not provide strong privacy guarantees.

## 8 Conclusion

This paper presents PrivPetal, a novel algorithm for synthesizing relational data with foreign keys under $(\epsilon, \delta)$-DP. PrivPetal enumerates possible permutations of tuples in permutation relations, and uses marginals queried from them to synthesize attributes sequentially. As these NPMs capture inter-attribute, intra-group, and inter-relational correlations, the generated synthetic data mimics the original database with high utility. Extensive experiments demonstrate the superiority of PrivPetal against state-of-the-art DP synthesis methods. For future work, we plan to explore how large language models (LLMs) can be utilized to enhance data synthesis under DP. For example, we might utilize a public pre-trained LLM to reduce the number of NPM queries since we can already obtain an educated guess for data correlations through the LLM, which promises a significant reduction of privacy cost.

## A  Proof of Theorem 3.1

By the definitions of $\mathbb{M}_{S',s}$ and NPMs, we have:

$$\mathbb{M}_{S',s}[x] = \sum_{t \in \mathbb{F}_s(R_I, R_H), t[S']=x} 1, \tag{9}$$

$$M_{S,s}[x] = \frac{1}{W_s} \sum_{t \in P_s(R_I, R_H), t[S]=x} 1. \tag{10}$$

Let $m$ be an injective mapping, from each letter $x$ to its order in the alphabet, i.e., $m(a) = 1, m(b) = 2, \ldots$. Let $S''$ be given by replacing each individual identifier $I_x$ in $S$ with $I_{m(x)}$.

Recall that $P_s(R_I, R_H)$ enumerates $s$-individual permutations, instead of $o$-individual permutations, when $s < o$. Let $o' = \min\{s, o\}$. Then, $P_s(R_I, R_H)$ contains all the $o'$-individual permutations. In addition, $\mathbb{F}_s(R_I, R_H)$ contains all the $s$-individual permutations.

Notice that the first $o'$-individuals in the $s$-individual permutations, are also $o'$-individual permutations, with each repeated by $(s - o')!$ times. As such, counting $S$ values from $P_s(R_I, R_H)$ is equivalent to counting $S''$ values from $\mathbb{F}_s(R_I, R_H)$, with a multiplier $\frac{1}{(s-o')!}$. By Eq. (10), we have:

$$M_{S,s}[x] = \frac{1}{W_s} \cdot \frac{1}{(s-o')!} \sum_{t \in \mathbb{F}_s(R_I, R_H), t[S'']=x} 1.$$

Substituting $W_s$ with Eq. (1), we have:

$$M_{S,s}[x] = \begin{cases} \frac{(s-o)!}{s!} \cdot \frac{1}{(s-o')!} \sum_{t \in \mathbb{F}_s(R_I, R_H), t[S'']=x} 1, & \text{if } s \geq o, \\ \frac{1}{s!} \cdot \frac{1}{(s-o')!} \sum_{t \in \mathbb{F}_s(R_I, R_H), t[S'']=x} 1, & \text{if } s < o, \end{cases}$$

$$= \frac{1}{s!} \sum_{t \in \mathbb{F}_s(R_I, R_H), t[S'']=x} 1. \tag{11}$$

Since both $S''$ and $S'$ can be obtained by an injective mapping from $S$, there exists an injective mapping $m'$ from integers to integers, such that we can convert $S''$ to $S'$ by applying $m'$ to the individual identifiers in $S''$. We can then move the individual columns in $\mathbb{F}_s(R_I, R_H)$, following this mapping $m'$, to create a new relation $\mathbb{F}'_s(R_I, R_H)$. Specifically, $\forall i, j \in [N]$ if $m'(i) = j$, then all $I_i$ columns in $\mathbb{F}_s(R_I, R_H)$ will be the $I_j$ columns in $\mathbb{F}'_s(R_I, R_H)$. As such, counting $S''$ values in $\mathbb{F}_s(R_I, R_H)$ is equivalent to counting $S'$ values in $\mathbb{F}'_s(R_I, R_H)$. By Eq. (11), we have

$$M_{S,s}[x] = \frac{1}{s!} \sum_{t \in \mathbb{F}'_s(R_I, R_H), t[S']=x} 1.$$

Noticing that moving individuals in all permutations results in the same set of permutations, $\mathbb{F}_s(R_I, R_H)$ is the same as $\mathbb{F}'_s(R_I, R_H)$. As such, we have:

$$M_{S,s}[x] = \frac{1}{s!} \sum_{t \in \mathbb{F}_s(R_I, R_H), t[S']=x} 1.$$

Combining it with Eq. (9), we conclude:

$$M_{S,s}[x] = \frac{1}{s!} \mathbb{M}_{S',s}[x].$$

## B  Proof of Theorem 5.3

We first introduce a lemma regarding the $L_2$ sensitivity when neighbor databases differ by at most $\tau$ tuples in a relation. Then, we quantify the privacy cost of PrivMRF.

LEMMA B.1. *Let $f$ be a function that maps a relation to a real vector. Let $\Delta(f)$ denote its $L_2$ sensitivity when the neighbor database is defined by removing/adding a tuple from a single-relation database. Let $\Delta'(f)$ denote the $L_2$ sensitivity when the input relation differs by at most $\tau$ tuples in the neighbor databases. We have:*

$$\Delta'(f) = \tau \Delta(f).$$

PROOF. Let $R$ and $R'$ be two relations such that $R'$ differs from $R$ by at most $\tau$ tuples. We have:

$$\Delta'(f) = \max_{R,R'} \|f(R) - f(R')\|_2.$$

We can create a sequence of relation $R_1, R_2, \ldots, R_{\tau'+1}$ such that (i) $\tau' \leq \tau$, (ii) $\forall i < \tau' + 1$, $R_i$ and $R_{i+1}$ differ by at most one tuple, and (iii) $R_1 = R$ and $R_{\tau'+1} = R'$. By the triangle inequality of $L_2$ norm, we have:

$$\Delta'(f) \leq \max_{R,R'} \sum_{i<\tau'+1} \|f(R_i) - f(R_{i+1})\|_2$$
$$\leq \sum_{i<\tau'+1} \Delta(f)$$
$$\leq \tau \Delta(f).$$

□

In PrivMRF, the input database contains a single relation, and neighbor databases are defined by removing/adding a tuple from that relation. In our method, the number of tuples in relation $R'$ can change is $\tau$. The above lemma shows that the $L_2$ sensitivity of any query imposed by PrivMRF is $\tau$ times the original $L_2$ sensitivity. By definition, its privacy cost is proportional to its squared $L_2$ sensitivities. As such, the privacy cost of applying PrivMRF to $R'$ is $\tau^2 C_H$.

## C  Proof of Theorem 5.4

As Algorithm 2 is a subroutine of Algorithm 1, we begin by bounding the $L_2$ sensitivities of the queries used by Algorithm 2 and Algorithm 1. Then, we quantify the privacy costs.

LEMMA C.1. *The $L_2$ sensitivity of querying $M_{S,s}$ for some given attribute set $S$ and all possible group sizes $s$ as a whole is at most $\tau$.*

PROOF. For any neighbor database, we denote the number of removed/added tuples of group size $s$ in $R_H$ (or $R'$) as $\tau_s$. Then, we have $\sum_s \tau_s \leq \tau$, and $P_s(R_I, R_H)$ differs by at most $W_s \tau_s$ tuples in the neighbor database. Let $M'_{S,s}$ be the corresponding NPM counted from the neighbor database. Since NPMs are contingency tables weighted by $\frac{1}{W_s}$, we have:

$$\|M_{S,s} - M'_{S,s}\|_1 \leq \frac{1}{W_s} \cdot W_s \tau_s$$
$$= \tau_s \tag{12}$$

Then, we have:

$$\|M_{S,s} - M'_{S,s}\|_2 \leq \|M_{S,s} - M'_{S,s}\|_1$$
$$\leq \tau_s.$$

Then, the $L_2$ sensitivity of NPMs for all group sizes is bounded by:

$$\sqrt{\sum_s \|M_{S,s} - M'_{S,s}\|_2^2} \leq \sqrt{\sum_s \tau_s^2} \leq \sum_s \tau_s \leq \tau.$$

□

LEMMA C.2. *The $L_2$ sensitivity of the h-score is at most $\tau$.*

PROOF. Let $h(S)$ and $h'(S)$ be two h-scores queried from two neighbor databases, respectively. Let $M'_{S,s}$ be the NPM counted from the neighbor database. By definition, we have:

$$\|h(S) - h'(S)\|_2 = \left\|\sum_{s:s\geq i}\|M_{S,s} - p_{S,s}\|_1 - \sum_{s:s\geq i}\|M'_{S,s} - p_{S,s}\|_1\right\|_2$$

$$= \left|\sum_{s:s\geq i}\|M_{S,s} - p_{S,s}\|_1 - \sum_{s:s\geq i}\|M'_{S,s} - p_{S,s}\|_1\right|$$

$$= \left|\sum_{s:s\geq i}\left(\|M_{S,s} - p_{S,s}\|_1 - \|M'_{S,s} - p_{S,s}\|_1\right)\right|$$

$$\leq \sum_{s:s\geq i}\left|\|M_{S,s} - p_{S,s}\|_1 - \|M'_{S,s} - p_{S,s}\|_1\right|$$

$$\leq \sum_{s:s\geq i}\|M_{S,s} - p_{S,s} - (M'_{S,s} - p_{S,s})\|_1$$

$$= \sum_{s:s\geq i}\|M_{S,s} - M'_{S,s}\|_1,$$

The second equation is because h-scores are scalars. Similar to the proof of Lemma C.1, by Eq. (12), we then have:

$$\|h(S) - h'(S)\|_2 \leq \sum_{s:s\geq i} \tau_s$$

$$\leq \tau.$$

□

LEMMA C.3. *The $L_2$ sensitivity of the R-score is at most $2\tau$.*

PROOF. For any attributes $A_1, A_2 \in \mathcal{A}_N$, and any possible group size $s$, we define:

$$R_s(A_1, A_2) = \|C_{\{A_1,A_2\},s} - \frac{1}{n_s W_s}C_{\{A_1\},s} \otimes C_{\{A_2\},s}\|_1,$$

where $C_{\{A_1,A_2\},s}, C_{\{A_1\},s}$, and $C_{\{A_2\},s}$ are marginals counted from $P_s(R, R')$. We similarly define $R'_s(A_1, A_2)$ as the corresponding value obtained from the neighbor database. Let $R'(A_1, A_2)$ be the R-score obtained from the neighbor database. By Eq. (3), we have:

$$\|R(A_1, A_2) - R'(A_1, A_2)\|_2 = |R(A_1, A_2) - R'(A_1, A_2)|$$

$$\leq \sum_s \frac{1}{W_s}\left|R_s(A_1, A_2) - R'_s(A_1, A_2)\right|.$$

Since $n_s W_s$ is the total number of tuples in $P_s(R, R')$, $R_s(A_1, A_2)$ is exactly the same as the R-score in [41]. By Lemma 4 of [6], the $L_2$ sensitivity of this R-score is 2 when neighbor databases differ by at most one tuple.

Let $\tau_s$ be the number of tuples of group size $s$ removed/added in $R'$ in our neighbor database. Then, our neighbor databases differ by at most $\tau_s W_s$ tuples in $P_s(R, R')$. The difference between $R_s$ and $R'_s$ is

correspondingly scaled by $\tau_s W_s$, similar to the proof in Lemma B.1. As such, we have:

$$\|R(A_1, A_2) - R'(A_1, A_2)\|_2 \leq \sum_s \frac{1}{W_s}\tau_s W_s \cdot 2$$

$$\leq 2\tau.$$

□

Next, we quantify the privacy cost of Algorithm 2 with the above lemmas.

LEMMA C.4. *The privacy cost of applying Algorithm 2 is*

$$C_{MRF} = \tau^2 \frac{kT_2}{\sigma_h^2} + \tau^2 \frac{T_2}{\sigma_M^2}. \quad (13)$$

PROOF. Its queries and the corresponding privacy costs are listed as follows:

(1) **h-scores**: Algorithm 2 samples $k$ attribute sets at each of its $T_2$ iterations and queries their h-scores in Line 11. The total number of h-scores is $kT_2$. By Lemma C.2, the $L_2$ sensitivity of h-score is $\tau$. The privacy cost of these h-scores is $\tau^2 \frac{kT_2}{\sigma_h^2}$.

(2) **NPMs**: Algorithm 2 queries NPMs for an attribute set $S$ at each of its $T_2$ iterations in Line 13. The total number of these queries is $T_2$. By Lemma C.1, the $L_2$ sensitivity of each query is $\tau$. The privacy cost of these queries is $\tau^2 \frac{T_2}{\sigma_M^2}$.

Then, the privacy cost of applying Algorithm 2 is the summation of these terms. □

Now, we are ready to quantify the privacy cost of Algorithm 1 with Lines 1-2 removed. Its queries and the corresponding privacy costs are listed as follows:

(1) **R-scores**: It queries the R-scores of for all attribute pairs in Line 3. By the definition of Eq. (3), R-scores are calculated from NPMs. As NPMs can be derived from other NPMs, R-scores are similarly derived from other R-scores, by mapping individual identifiers. Therefore, it suffices to query the basic R-scores, and derive all other R-scores accordingly. Each R-score query has a $L_2$ sensitivity of $2\tau$. The number of such basic R-score queries is list as follows.
   (a) A total of $|\mathcal{A}_H||\mathcal{A}_I|$ queries for $R(A_1, A_2), \forall A_1 \in \mathcal{A}_H, A_2 \in \mathcal{A}_{I_1}$.
   (b) A total of $|\mathcal{A}_H||\mathcal{A}_H - 1|$ queries for $R(A_1, A_2), \forall A_1, A_2 \in \mathcal{A}_H$ and $A_1 \neq A_2$.
   (c) A total of $|\mathcal{A}_I||\mathcal{A}_I - 1|$ queries for $R(A_1, A_2), \forall A_1, A_2 \in \mathcal{A}_{I_1}$ and $A_1 \neq A_2$.
   (d) A total of $\frac{|\mathcal{A}_I||\mathcal{A}_I+1|}{2}$ queries for $R(I_1.A_i, I_2.A_j), \forall I_1.A_i \in \mathcal{A}_{I_1}, I_2.A_j \in \mathcal{A}_{I_2}$, and $i \leq j$.
   
   Then, the total number of all these queries is the summation:

$$|\mathcal{A}_H||\mathcal{A}_I| + |\mathcal{A}_H||\mathcal{A}_H - 1| + |\mathcal{A}_I||\mathcal{A}_I - 1| + \frac{|\mathcal{A}_I||\mathcal{A}_I + 1|}{2}$$

$$= \frac{1}{2}(|\mathcal{A}_H|^2 + 2|\mathcal{A}_H||\mathcal{A}_I| + 2|\mathcal{A}_I|^2 - |\mathcal{A}_H|).$$

By Lemma C.3, the $L_2$ sensitivity of R-score is $2\tau$. The privacy cost is $2\tau^2 \frac{|\mathcal{A}_H|^2 + 2|\mathcal{A}_H||\mathcal{A}_I| + 2|\mathcal{A}_I|^2 - |\mathcal{A}_H|}{\sigma_R^2}$.

**Algorithm 5:** Initialization of the Noisy NPM Set

**Input:** Referencing relation $R_I$, referenced relation $R_H$, privacy cost $C$, noisy R-scores $\widetilde{R(\cdot,\cdot)}$, noisy group sizes $\{\widetilde{n_s}\}$.
**Output:** $\widetilde{\mathcal{M}_{\text{all}}}$.

// Inter-attribute correlations
1 Invoke PrivMRF to synthesize $R_H$, producing $\widetilde{R_H}$, with privacy cost $\frac{C}{4}$;
2 Decompose the noisy marginals queried by PrivMRF into NPMs;
3 Invoke PrivMRF to synthesize $\pi_{\mathcal{A}_{I_1}}(P_s(R_I, R_H))$ for each group size $s \leq N$, each with privacy cost $\frac{C}{4}$;
4 Roll up all the noisy NPMs obtained by PrivMRF to smaller NPMs, and store all NPMs in $\widetilde{\mathcal{M}_{\text{all}}}$;
5 Calculate noise scales $\sigma_{\text{inter}}, \sigma_{\text{intra}}$, each based on a privacy cost of $\frac{C}{4}$;

// Inter-relational correlations
6 Let $C_{\text{inter}}$ be a set such that each $S \in C_{\text{inter}}$ (i) is a subset of $\mathcal{A}_H \cup \mathcal{A}_I$, (ii) contains at least one attribute from $\mathcal{A}_H$ and one attribute from $\mathcal{A}_I$, and (iii) satisfies $\lambda$-usefulness [41];
7 **for** $A \in \mathcal{A}_H \cup \mathcal{A}_{I_1}$ **do**
8     Select $S \in C_{\text{inter}}$ for attribute $A$ with CFS [16];
9     $\widetilde{\mathcal{M}_{\text{all}}} \leftarrow$ Algorithm 3$(S, \widetilde{\mathcal{M}_{\text{all}}}, \{\widetilde{n_s}\}, \sigma_{\text{inter}})$;

// Intra-group correlations
10 **for** $i = 2$ **to** $o$ **do**
11     Let $C_{\text{intra}}$ be a set such that each $S \in C_{\text{intra}}$ (i) is a subset of $\bigcup_{j=1}^{i} \mathcal{A}_{I_j}$, (ii) contains all $i$ different individual identifiers, and (iii) satisfies $\lambda$-usefulness;
12     **for** $A \in \mathcal{A}_{I_i}$ **do**
13         Select $S \in C_{\text{intra}}$ for attribute $A$ with CFS;
14         $\widetilde{\mathcal{M}_{\text{all}}} \leftarrow$ Algorithm 3$(S, \widetilde{\mathcal{M}_{\text{all}}}, \{\widetilde{n_s}\}, \sigma_{\text{intra}})$;

15 **return** $\widetilde{\mathcal{M}_{\text{all}}}$;

(2) **Number of Flattened Tuples for Each Group Size**: Algorithm 1 queries the number $n_s$ of flattened tuples for each group size $s < N$ in the FR in Line 4. Such counting queries as a whole have a $L_2$ sensitivity of $\tau$, because removing or adding $\tau$ tuples in $R_H$ (or $R'$) can change all counts $n_s$ by at most $\tau$. The privacy cost is $\frac{\tau^2}{\sigma_n^2}$.

(3) **MRF Construction**: It invokes Algorithm 2 to construct MRFs for each individual attribute in Line 7. The total number of invocations is $N|\mathcal{A}_I|$. The total privacy cost for these invocations is $N|\mathcal{A}_I|C_{\text{MRF}}$.

Therefore, the total privacy cost is:

$$C_{\text{one-FK}} = 2\tau^2 \frac{|\mathcal{A}_I|^2 + 2|\mathcal{A}_H||\mathcal{A}_I| + 2|\mathcal{A}_I|^2 - |\mathcal{A}_H|}{\sigma_R^2}$$
$$+ \frac{\tau^2}{\sigma_n^2} + \tau^2 T_2 N |\mathcal{A}_I| \left( \frac{k}{\sigma_h^2} + \frac{1}{\sigma_M^2} \right).$$

## D  Initialization of the noisy NPM set $\widetilde{\mathcal{M}_{\text{all}}}$

This section presents a general approach for initializing $\widetilde{\mathcal{M}_{\text{all}}}$ in Algorithm 1. The initialized $\widetilde{\mathcal{M}_{\text{all}}}$ is subsequently sent to Algorithm 2 to construct MRFs for synthesizing individual attributes iteratively in Lines 6-9.

Recall that our Algorithm 2 mostly uses the NPMs from the given $\widetilde{\mathcal{M}_{\text{all}}}$ to construct MRFs, and only selects $T_2 = 1$ new NPM. The quality of the resulting MRFs largely depends on whether $\widetilde{\mathcal{M}_{\text{all}}}$ contains the most correlated NPMs. This problem is particularly important during the initial iterations of individual attribute synthesis, as $\widetilde{\mathcal{M}_{\text{all}}}$ may not yet include sufficient newly added NPMs and primarily depends on the initial ones.

To address this, our goal is to systematically initialize $\widetilde{\mathcal{M}_{\text{all}}}$ to capture three types of correlations: inter-attribute, intra-group, and inter-relational. The pseudocode is provided in Algorithm 5. Among the inputs, privacy cost $C$ is used for querying NPMs to initialize $\widetilde{\mathcal{M}_{\text{all}}}$. We include a discussion of this privacy cost at the end of this section.

**Inter-Attribute Correlations.** Similar to Lines 1-2 in Algorithm 1, Algorithm 5 applies a modified version of PrivMRF to $R_H$, with a privacy cost of $\frac{C}{4}$ (Line 1), and decomposes the queried marginals into NPMs (Line 2).

In addition, Algorithm 5 applies another modified version of PrivMRF to the $I_1$ part of each PR, i.e., $\pi_{\mathcal{A}_{I_1}}(P_s(R_I, R_H))$ for each group size $s < N$, to capture the inter-attribute correlations in $R_I$ (Line 3). Specifically, we simply change every marginal used by PrivMRF to the corresponding NPM, i.e., weight it with $\frac{1}{W_s}$. Algorithm 5 then rolls up the NPMs obtained in Lines 1 and 3 to smaller NPMs, and stores all these NPMs in $\widetilde{\mathcal{M}_{\text{all}}}$ (Line 4). The privacy cost of Line 3 is $\frac{C}{4}$ by the parallel composition of DP, since data of different group sizes are disjoint.

**Inter-Relational Correlations.** Algorithm 5 first calculates the noise scales $\sigma_{\text{inter}}$ and $\sigma_{\text{intra}}$ in Line 5 (explained later). Then, it constructs a candidate set $C_{\text{inter}}$ such that each $S \in C_{\text{inter}}$ (i) is a subset of $\mathcal{A}_H \cup \mathcal{A}_I$, (ii) contains at least one attribute from $\mathcal{A}_H$ and one attribute from $\mathcal{A}_I$, and (iii) satisfies $\lambda$-usefulness [41] (Line 6). The $\lambda$-usefulness ensures that each NPM $M_{S,s}$ remains high utility after the injection of Gaussian noise, and is defined as:

$$\frac{\widetilde{n}}{\sum_s |M_{S,s}|} \geq \lambda \sqrt{\frac{2}{\pi}} \sigma_{\text{inter}},$$

where the right side is the product of a hyperparameter $\lambda$ and the mean absolute deviation of Gaussian noise with scale $\sigma_{\text{inter}}$. This ensures that the mean ratio of NPM counts to noise level is at least $\lambda$, which is set to 6 as recommended by previous work [6].

Then, Algorithm 5 applies the *correlation-based feature selector (CFS)* [16] to select one NPM capturing inter-relational correlation, for each attribute $A \in \mathcal{A}_H \cup \mathcal{A}_{I_1}$ (Lines 7-9). Using noisy R-scores to measure correlations, for each candidate $S \in C_{\text{inter}}$, the CFS metric is defined as:

$$\rho(A, S) = \frac{\sum_{A' \in S \setminus \{A\}} \widetilde{R(A, A')}}{\sqrt{|S| + \sum_{A' \in S \setminus \{A\}} \sum_{A'' \in S \setminus \{A, A'\}} \widetilde{R(A', A'')}}}. \quad (14)$$

This metric rewards the correlations between the target attribute $A$ and the remaining attributes in $S \setminus \{A\}$ while penalizing the correlations among the attributes in $S \setminus \{A\}$. The penalty term aims to minimize the redundant information contained by the attributes in $S$. By selecting one $S$ with the highest $\rho(A, S)$ for each $A \in \mathcal{A}_H \cup \mathcal{A}_I$, Algorithm 5 ensures the inclusion of at least one NPM that captures strong inter-relational correlations for each $A$.

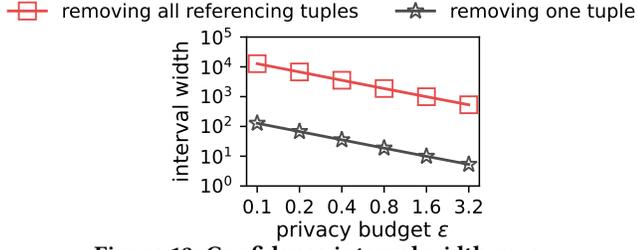

Figure 13: Confidence interval width vs. $\epsilon$.

**Intra-Group Correlations**. Algorithm 5 selects NPMs capturing intra-group correlations for each order $i = 2, \ldots, o$, i.e., NPMs containing $i = 2, \ldots, o$ different individual identifiers (Lines 10-14). It selects NPMs for each order $i$ separately to ensure that both low-order correlations and high-order correlations can be captured. Similar to selecting NPMs for inter-relational correlations, for each order $i$, it constructs a candidate set $C_{\text{intra}}$ such that each $S \in C_{\text{intra}}$ (i) is a subset of $\bigcup_{j=1}^{i} \mathcal{A}_{I_j}$, (ii) contains all $i$ different individual identifiers, and (iii) satisfies $\lambda$-usefulness (Line 11).

Then, for each attribute $A \in \mathcal{A}_{I_i}$, Algorithm 5 applies CFS to select one NPM capturing intra-group correlation (Line 13), and queries the corresponding NPM (Line 14). Intuitively, the selected $S$ contains intra-group correlations that can be used for synthesizing each $A$.

**Privacy Analysis**. Algorithm 5 spends the privacy cost as follows. Recall that PrivMRF also utilizes analytic Gaussian mechanism, Algorithm 5 simply provides the required privacy cost $\frac{C}{4}$ to the invocations of PrivMRF in Lines 1 and 3, to run PrivMRF for inter-attribute correlations.

Since Algorithm 5 uses the given noisy R-scores to select NPMs for inter-relational and intra-group correlations, the selection process does not incur any privacy cost. Thus, it can use all the remaining privacy costs to query NPMs. Specifically, it splits the remaining $\frac{C}{2}$ evenly for inter-relational and intra-group correlations, and thus, calculates the noise scales as:

$$(|\mathcal{A}_H| + |\mathcal{A}_I|)\frac{\tau^2}{\sigma_{\text{inter}}^2} = \frac{C}{4},$$

$$(o-1)|\mathcal{A}_I|\frac{\tau^2}{\sigma_{\text{intra}}^2} = \frac{C}{4}.$$

In summary, (i) the invocation of PrivMRF for household attributes spends $\frac{C}{4}$, (ii) the invocations of PrivMRF for individual attributes spend $\frac{C}{4}$, (iii) querying NPMs for inter-relational correlations spends $\frac{C}{4}$, and (iv) querying NPMs for intra-group correlations spends $\frac{C}{4}$. The total privacy cost is $C$, as designated in the inputs of Algorithm 5.

Finally, Algorithm 5 returns $\widetilde{\mathcal{M}_{\text{all}}}$ as the output, which stores all the obtained NPMs. Algorithm 1 can replace its Lines 1-2 with Algorithm 5 to initialize $\widetilde{\mathcal{M}_{\text{all}}}$, and accordingly, incurs a larger privacy cost.

THEOREM D.1. *The privacy cost of applying Algorithm 1 to R and R′, with Lines 1-2 replaced by Algorithm 5, is:*

$$C_{\text{one-FK}} = C + 2\tau^2 \frac{|\mathcal{A}_H|^2 + 2|\mathcal{A}_H||\mathcal{A}_I| + 2|\mathcal{A}_I|^2 - |\mathcal{A}_H|}{\sigma_R^2}$$
$$+ \frac{\tau^2}{\sigma_n^2} + \tau^2 T_2 N |\mathcal{A}_I| \left(\frac{k}{\sigma_h^2} + \frac{1}{\sigma_M^2}\right).$$

In general, Algorithm 4 utilizes Algorithm 1, with Lines 1-2 replaced by Algorithm 5, to process each foreign key. Since the privacy cost $C$ is used for querying NPMs to capture basic correlations, and can be used to construct all MRFs, we typically set $C = \frac{1}{2}C_{\text{one-FK}}$ to ensure a good initialization.

## E Comparative Study of Different Definitions of Neighboring Databases

As explained in Section 2.1, in our problem setting, a neighboring database is obtained by removing one tuple from the primary private relation and all other tuples that depend on it. In contrast, Ghazi et al. [14] adopt a different notion of neighboring databases, which is defined as follows:

*Definition E.1 (Neighboring Database Without Enforcing Foreign Key Constraints).* Two databases are considered neighboring if and only if one can be derived from the other by removing exactly one tuple from any relation in the database.

Both definitions are widely used in the DP literature, but our privacy notion provides stronger protection. For example, consider the users and orders tables in the Instacart dataset. Suppose that an adversary knows that a specific user, Alice, has the largest number of orders among all users, and the adversary aims to infer the exact number of Alice's orders, denoted by $\alpha$. Intuitively, a stricter privacy guarantee should makes it more difficult for the adversary to estimate $\alpha$ accurately.

Suppose that we use the analytic Gaussian mechanism to release the maximum number of orders any user actually has (i.e., $\alpha$ in this case). Let $M$ be the largest number of orders that a user can have. Under our definition of neighboring databases, the $L_2$ sensitivity of $\alpha$, denoted as $\Delta_1$, is equal to $M$. This is because, in the worst case, (i) one user may have $M$ orders while all other users have none, and (ii) removing the user with $M$ orders could reduce the maximum from $M$ to 0. Accordingly, the release value is computed as:

$$\widetilde{\alpha_1} = \alpha + \mathcal{N}(0, \sigma_1),$$

where

$$\gamma = \sqrt{\left(\frac{\Delta_1}{\sigma_1}\right)^2},$$
$$\Delta_1 = M,$$

and $\gamma$ is determined by the desired $(\epsilon, \delta)$-DP guarantee. We can compute $\gamma$ using Eq. (4), and then solve for $\sigma_1$ to obtain $\widetilde{\alpha_1}$.

If we instead adopt Definition E.1, the $L_2$ sensitivity of $\alpha$ is 1, as removing any single tuple from the database changes the maximum number of orders by at most 1. In this case, the released value is:

$$\widetilde{\alpha_2} = \alpha + \mathcal{N}(0, \sigma_2),$$

where

$$\gamma = \sqrt{\left(\frac{\Delta_2}{\sigma_2}\right)^2},$$
$$\Delta_2 = 1.$$

The ratio of the noise scales between these two definitions is $\frac{M}{1}$, indicating that our definition can provide significantly stronger privacy protection when $M$ is large.

To validate the above analysis, we conduct experiments using the Instacart dataset, where a user can have at most 100 orders (i.e., $M = 100$). Figure 13 reports the width of the adversary's 95% confidence interval for each definition when inferring $\alpha$. (This interval is derived from the 95% confidence interval of the Gaussian noise added to $\alpha$.) Results based on our definition are labeled as "removing all referencing tuples", while results based on Definition E.1 are labeled as "removing one tuple". A narrower confidence interval indicates a more accurate estimate of the true value. Under our definition of neighboring databases, the adversary's estimate for $\alpha$ is much less precise compared to the estimate obtained using Definition E.1. In particular, when $\epsilon = 3.2$, the width of the confidence interval pertinent to Definition E.1 is about 5.33. That is, the adversary can infer with 95% confidence that the true number of Alice's orders lies within an interval of width less than 5.33. In contrast, when our definition is adopted, the width of the confidence interval is around 533, and hence, Alice's privacy is much better protected. These results support our analysis that our privacy notion can offer a stronger guarantee.